\documentclass[11pt]{article}
\usepackage{graphicx}
\usepackage{amssymb}
\usepackage{amsmath, amsthm}
\newcommand{\eq}{\!=\!}
\newcommand{\ee}{\end{equation}}
\newcommand{\word}[1]{\,\,\mbox{#1}\,\,}
\newcommand{\reff}[1]{(\ref{#1})}
\newcommand{\beq}{\begin{equation}}
\newcommand{\eeq}[1]{\label{#1}\end{equation}}
\newcommand{\beg}{\begin{equation*}}
\newcommand{\eeg}{\end{equation*}}

\newcommand{\expo}[1]{\mbox{e}^{#1}}

\newcommand{\less}{\!<\!}

\newcommand{\sumprime}{\sideset{}{'}\sum}

\newcommand{\bsplit}{\begin{split}}
\newcommand{\esplit}{\end{split}}

\begin{document}
\def\theequation{\arabic{section}.\arabic{equation}}
\begin{titlepage}
\title{Cancellation of nonrenormalizable hypersurface divergences and the $d$-dimensional Casimir piston}
\author{\thanks{Email: aedery@ubishops.ca} Ariel Edery and Ilana
MacDonald\\
{\small \it Physics Department, Bishop's University}\\
{\small \it 2600 College Street, Sherbrooke, Qu\'{e}bec, Canada
J1M~0C8}}

\date{} \maketitle
\thispagestyle{empty} \vspace*{1truecm}

\begin{abstract}
\noindent Using a multidimensional cut-off technique, we obtain
expressions for the cut-off dependent part of the vacuum energy for
parallelepiped geometries in any spatial dimension $d$. The cut-off
part yields nonrenormalizable hypersurface divergences and we show
explicitly that they cancel in the Casimir piston scenario in all
dimensions. We obtain two different expressions for the
$d$-dimensional Casimir force on the piston where one expression is
more convenient to use when the plate separation $a$ is large and
the other when $a$ is small (a useful $a \to 1/a$ duality). The
Casimir force on the piston is found to be attractive (negative) for
any dimension $d$. We apply the $d$-dimensional formulas (both
expressions) to the two and three-dimensional Casimir piston with
Neumann boundary conditions. The 3D Neumann results are in numerical
agreement with those recently derived in arXiv:0705.0139 using an
optical path technique providing an independent confirmation of our
multidimensional approach.  We limit our study to massless scalar
fields.

\end{abstract}
\vspace*{1truecm} %
\noindent
\begin{center}
\end{center}
\setcounter{page}{1}
\end{titlepage}

\def\theequation{\arabic{section}.\arabic{equation}}


\setcounter{page}{2}

\section{Introduction}
\setcounter{equation}{0}

With idealized boundary conditions, such as Dirichlet and Neumann
boundary conditions, vacuum energy calculations have cut-off
dependent terms that diverge in the limit as the cut-off $\Lambda$
tends to infinity. These divergences can be classified as either
bulk (volume) divergences or lower-dimensional surface divergences
(for simplicity, all divergences besides the volume divergence will
be classified as hypersurface or simply surface divergences
regardless of the dimension). The volume divergence poses no
problems since the Casimir energy, by definition, is obtained after
subtracting out the volume divergence from the vacuum energy. In
other words, the volume divergence is renormalizable and can be set
to zero by simply adding a constant counterterm to the Hamiltonian.
In contrast, the surface divergences are nonrenormalizable. It is
tempting to throw out the surface divergence as an artifact of
idealized boundary conditions and retain the finite part as the true
Casimir energy as is often done in the literature (see references in
\cite{Bordagreport}). However, this is not a physically valid
renormalization procedure. It has been shown that these surface
divergences cannot be removed via renormalization of any physical
parameters of the theory \cite{Weigel,Barton2,Barton3}. In the zeta-
function regularization technique, these surface divergences do not
appear because in effect they are renormalized to zero\footnote{Like
dimensional regularization, zeta-function regularization goes beyond
pure regularization and does some renormalization.}. In special
cases, like the parallel plate geometry with infinite plates, this
is not an issue because in the limit as $\Lambda\to\infty$ the
Casimir force is finite. However, in any realistic situation where
the plates are of finite size, the Casimir force diverges in the
limit $\Lambda\to \infty$.

Unambiguous Casimir calculations can be carried out with idealized
boundary conditions in the apparatus called the Casimir piston. A
few years ago, Cavalcanti \cite{Cavalcanti} showed for the case of a
two-dimensional (2+1) massless scalar field confined to a
rectangular region with Dirichlet boundary conditions that the
Casimir piston can resolve the issue of nonrenormalizable surface
divergences that appear in Casimir calculations. A Casimir piston
contains an interior and an exterior region and Cavalcanti showed
explicitly that the surface cut-off terms of the interior and
exterior regions canceled. He also showed that the Casimir force on
the piston is always negative regardless of the ratios of the two
sides of the rectangular region. This is in contrast to calculations
that can yield positive Casimir forces in rectangular geometries
when the surface cut-off terms are thrown out and no exterior region
is considered (see references in \cite{Bordagreport}).

In this paper, we use a multidimensional cut-off technique
\cite{Ariel} to obtain exact expressions for the cut-off dependent
($\Lambda$-dependent) part of the Casimir energy for a
$d$-dimensional parallelepiped region. In the limit
$\Lambda\!\to\!\infty$, these yield nonrenormalizable hypersurface
divergences and we show explicitly that they cancel out in the
Casimir piston scenario for any dimension. We then derive exact
expressions for the Casimir force on the piston in any dimension $d$
and use the invariance of the vacuum energy under permutations of
lengths to derive an alternative expression. When the plate
separation $a$ is large, an otherwise long computation using the
first expression becomes trivial using the alternative expression
and vice versa when $a$ is small (there is a useful $a\to 1/a$
duality). As in two and three dimensions, the Casimir force on the
piston is attractive (negative) in any spatial dimension $d$.

For the three-dimensional Casimir piston with massless scalar fields
obeying Dirichlet and Neumann boundary conditions approximate
results were first obtained in \cite{Kardar} for small plate
separation. Exact results for arbitrary plate separation were then
obtained for the Dirichlet case in \cite{Ariel2}. Exact results for
the 3D Neumann (as well as Dirichlet) case were recently obtained
via an optical path technique \cite{Hertzberg}. In \cite{Hertzberg},
arbitrary cross sections, temperature and free energy were also
studied. We apply our $d$-dimensional formulas (both expressions) to
the 2D and 3D Neumann cases. The first 3D expression looks similar
in form to the one recently derived in \cite{Hertzberg} and is in
numerical agreement with it. The alternative 3D expression converges
quickly when $a$ is large and though it is quite different in form
compared to the first expression or the one found in
\cite{Hertzberg} it is in numerical agreement with both of them. The
2D Neumann results are new and bring a completeness to the original
work of Cavalcanti \cite{Cavalcanti} where the 2D Dirichlet case was
considered. Before discussing the literature on Casimir pistons for
the electromagnetic case, it is worth noting that the use of
massless scalar fields in Casimir studies goes beyond theoretical
interest and has direct application to physical systems such as
Bose-Einstein condensates \cite{Ariel3,Pomeau,Visser}.
Higher-dimensional scalar field Casimir calculations have also been
carried out in the context of 6D supergravity theories
\cite{Burgess}.

For perfect-conductor conditions, the Casimir piston for the
electromagnetic field in a three-dimensional rectangular cavity was
studied in \cite{Kardar} and the Casimir force on the piston was
found to be attractive in contrast to results without exterior
region where the force could be positive. This was then generalized
further in Refs. \cite{Marachevsky1}-\cite{Hertzberg} where the
temperature and free energy dependence was studied. It was shown in
\cite{Klich} that the Casimir force between two bodies related by
reflection is always attractive, independent of the exact form of
the bodies or dielectric properties and this was generalized further
in \cite{Bachas}. It has also been shown that Casimir piston
scenarios can yield repulsive forces. The Casimir piston for a
weakly reflecting dielectric was considered in \cite{Barton} and it
was shown that though attraction occurred for small plate
separation, this could switch to repulsion for sufficiently large
separation. However, the force remained attractive for all plate
separations if the material was thick enough, in agreement with the
results in \cite{Kardar}. Two preprints \cite{Fulling,Zhai} also
discuss scenarios where repulsive Casimir forces in pistons can be
achieved. Recently, two independent groups have developed techniques
for calculating Casimir forces between arbitrary compact objects and
have applied the results to the case of two spherical bodies at a
distance \cite{Emig,Klich2}.

\section{Cancellation of hypersurface divergences in the $d$-dimensional Casimir
piston} \setcounter{equation}{0}

The expression for the vacuum energy $\tilde{E}$ regularized using a
multidimensional cut-off technique \cite{Ariel} divides naturally
into two parts: a finite part $E_0$ and a cut-off dependent part
$E(\Lambda)$ which diverges as the cut-off $\Lambda$ tends to
infinity. The expressions for $E(\Lambda)$ and $E_0$ are derived in
appendix A and are written in a compact fashion with the help of the
$ordered$ symbol $\xi^d_{k_1...k_j}$ which was introduced in
\cite{Ariel2} and is defined below after the expressions for
$E(\Lambda)$ are written. Our goal in this section is to show that
for the Casimir piston scenario, the hypersurface divergences of the
interior and exterior regions of the piston cancel out for any
dimension $d$. By ``cancel out" we do not mean that the cut-off
dependent part of the Casimir energy is zero but that it is
independent of the plate separation $a$ so that the Casimir {\it
force} on the piston has no cut-off dependence. In the next section
we focus on the finite part $E_0$ and obtain explicit expressions
for the Casimir force on the piston in any dimension. We work in
units where $\hbar\!=\!c\!=\!1$.

The regularized vacuum energy for massless scalar fields in a
$d$-dimensional box with sides of arbitrary lengths
$L_1,L_2,...,L_d$ obeying Dirichlet (D) boundary conditions is given
by  \reff{ED22} (as $\Lambda \to \infty$) \beq \tilde{E}_D =
E_{0_D}+E_D(\Lambda)\eeq{ED1} where $E_{0_D}$ is the finite part for
the Dirichlet case and $E_D(\Lambda)$ is the cut-off dependent part
given by \reff{EDCutoff}: \beq \begin{split} &E_D(\Lambda)\equiv
\dfrac{1}{2^{d+1}}\sum_{m=1}^d (-1
)^{d+m}\,m\,2^m\,\pi^{\frac{m+1}{2}}\,\Gamma(\tfrac{m+1}{2})\,\,\Lambda^{m+1}\,\,\xi^{\,d}_{\,k_1,..,
k_m}\prod_{i=1}^m\, L_{k_i}\,\,\,
\\&=\dfrac{1}{2^{d+1}}\sum_{m=1}^d (-1
)^{d+m}\,m\,2^m\,\pi^{\frac{m+1}{2}}\,\Gamma(\tfrac{m+1}{2})\,\,\Lambda^{m+1}\,\,\xi^{\,d}_{\,k_1,..,
k_m}\,(L_{k_1}\ldots L_{k_m}).
\end{split}\eeq{EDCutoff1} For the case of Neumann(N) boundary
conditions the regularized vacuum energy is given by
\reff{Neum-cut-off} (as $\Lambda \to \infty$)\beq \tilde{E}_N=
E_{0_N} + E_N(\Lambda)\eeq{EN1} where $E_{0_N}$ is the finite part
and $E_N(\Lambda)$  is the cut-off dependent part given by
\reff{ENCutoff}: \beq \begin{split} &E_N(\Lambda)\equiv
\dfrac{1}{2^{d+1}}\sum_{m=1}^d
m\,2^m\,\pi^{\frac{m+1}{2}}\,\Gamma(\tfrac{m+1}{2})\,\,\Lambda^{m+1}\,\,\xi^{\,d}_{\,k_1,..,
k_m}\prod_{i=1}^m\, L_{k_i}\,\,\,
\\&=\dfrac{1}{2^{d+1}}\sum_{m=1}^d m\,2^m\,\pi^{\frac{m+1}{2}}\,\Gamma(\tfrac{m+1}{2})\,\,\Lambda^{m+1}\,\,\xi^{\,d}_{\,k_1,..,
k_m}\,(L_{k_1}\ldots L_{k_m}).
\end{split}\eeq{ENCutoff1} In this section, the expressions for $E_{0_D}$ and
$E_{0_N}$ are not needed.

There is an implicit summation over the integers $k_i$ in
\reff{EDCutoff1} and \reff{ENCutoff1}. The ordered symbol
$\xi^{d}_{\,k_1,.., k_m}$ \cite{Ariel2} is defined by \beq
\xi^{\,d}_{\,k_1,.., k_m}=\begin{cases} 1&\word{if}
 k_1 \!<\!k_2\!<\! \ldots\! <k_m \,;\,1 \le k_m \le d\\ 0&
\word{otherwise}.
\end{cases}\eeq{ordered}
The ordered symbol ensures that the implicit sum over the $k_i$  is
over all distinct sets $\{k_1,\ldots, k_m\}$, where the $k_i$ are
integers that can run from $1$ to $d$ inclusively under the
constraint that $k_1\!<\!k_2\!<\!\cdots\!<k_m$. The superscript $d$
specifies the maximum value of $k_m$. For example, if $m=2$ and
$d=3$ then $\xi^{\,d}_{\,k_1,.., k_m}=\xi^{\,3}_{\,k_1, k_2}$ and
the non-zero terms are $\xi_{\,1,2}$ , $\xi_{\,1,3}$ and $
\xi_{\,2,3}$. This means the summation is over $\{k_1,k_2\}=(1,2),
(1,3)$ and $(2,3)$ so that $\xi^{\,3}_{\,k_1,k_2}\,L_{k_1}\,
L_{k_2}=L_1\,L_2+L_1\,L_3+L_2\,L_3$.

\subsection{Cancellation in three dimensions}

Before showing how the cut-off dependent hypersurface divergences in
the $d$-dimensional Casimir piston cancel, we consider the case of
three spatial dimensions first. Three dimensions allows us to make
the first non-trivial use of the $d$-dimensional cut-off expressions
\reff{EDCutoff1} and \reff{ENCutoff1} and to illustrate in a
transparent fashion how the cancellation occurs. This paves the way
to follow the cancellation in $d$-dimensions in the next subsection.
The cut-off expressions in three dimensions and their cancellation
in the piston scenario are in agreement with the work in
\cite{Kardar,Hertzberg} and this provides an independent
confirmation of our general formulas.

In three dimensions, the Dirichlet and Neumann cut-off expressions
$E_D(\Lambda)$ and $E_N(\Lambda)$, are obtained by substituting
$d=3$ in equations (\ref{EDCutoff1}) and (\ref{ENCutoff1}): \beq
\begin{split} E_D(\Lambda)&=\dfrac{1}{2^{4}}\sum_{m=1}^3 \,(-1
)^{3+m}
\,m\,2^m\,\pi^{\frac{m+1}{2}}\,\Gamma(\tfrac{m+1}{2})\,\Lambda^{m+1}\,\,\xi^{\,3}_{\,k_1,..,
k_m}\,(L_{k_1}\ldots L_{k_m})
\\&=\dfrac{\pi}{8}\Lambda^2(L_1+L_2+L_3)-\dfrac{\pi^2}{4}\Lambda^3(L_1\,L_2+L_1\,L_3+L_2\,L_3)
+\dfrac{3\pi^2}{2}\Lambda^4\,L_1\,L_2\,L_3 \end{split}\eeq{EDCut}
\beq \begin{split}E_N(\Lambda)&=\dfrac{1}{2^{4}}\sum_{m=1}^3
\,\xi^{\,3}_{\,k_1,.., k_m}\,(L_{k_1}\ldots
L_{k_m})\,m\,2^m\,\pi^{\frac{m+1}{2}}\,\Gamma(\tfrac{m+1}{2})\,\Lambda^{m+1}
\\&= \dfrac{\pi}{8}\Lambda^2(L_1+L_2+L_3)+\dfrac{\pi^2}{4}\Lambda^3(L_1\,L_2+L_1\,L_3+L_2\,L_3)
+\dfrac{3\pi^2}{2}\Lambda^4\,L_1\,L_2\,L_3\,.
\end{split}\eeq{ENCut}
Except for a trivial redefinition of $\Lambda \to \Lambda/\pi$, the
above cut-off expressions in three dimensions are in agreement with
those derived in \cite{Hertzberg}. The $\Lambda^4$ term appearing in
\reff{EDCut} and \reff{ENCut} is multiplied by the volume
$L_1\,L_2\,L_3$ of the box and represent the volume divergence of
the continuum. This volume term poses no divergence problems since
it must be subtracted to obtain the Casimir energy (defined as the
difference between the vacuum energy with boundaries and the bulk
vacuum energy of the continuum). In other words, it can be
renormalized to zero via a constant counterterm in the Hamiltonian.
The remaining $\Lambda^2$ and $\Lambda^3$ terms are proportional to
the perimeter and surface area respectively (we refer to both as
surface divergences for simplicity). In contrast to the volume
divergence, there is no physical justification for subtracting out
the surface divergences. In other words, they cannot be renormalized
to zero.

For the Casimir piston, the Casimir energy is obtained by adding the
vacuum energy of the interior region I and exterior region II (see
fig. 1). To obtain the cut-off dependence for the Casimir energy we
therefore add the cut-off terms (the surface divergences) in regions
I and II. Let the plate separation be $a$.
\begin{figure}
\begin{center}
\includegraphics[scale=0.9]{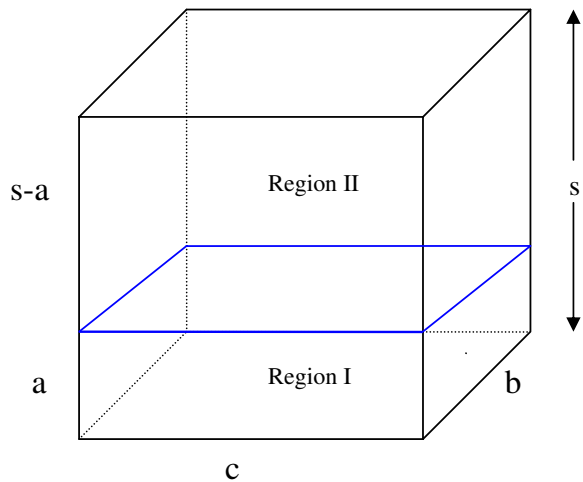}
\caption{Casimir piston in three dimensions.}
\end{center}
\end{figure}In region I, the three lengths are $L_1=a$, $L_2=b$
and $L_3=c$ whereas in region II the three lengths are $L_1=s-a$,
$L_2=b$ and $L_3=c$. Note that in region I, $L_1$ comes with $+\,a$
whereas in region II, $L_1$ comes with the opposite sign $-\,a$. For
Dirichlet boundary conditions we obtain \beq
\begin{split}E_{D_{piston}}(\Lambda)&= E_{D_1}(\Lambda)+
E_{D_2}(\Lambda)
\\&=\dfrac{\pi}{8}\Lambda^2(a+b+c)-\dfrac{\pi^2}{4}\Lambda^3(a\,b + a\,c
+b\,c)\\&+\dfrac{\pi}{8}\Lambda^2\big(s-a+b+c\big)-\dfrac{\pi^2}{4}\Lambda^3\big((s-a)\,b
+ (s-a)\,c
+b\,c\big)\\&=\dfrac{\pi}{8}\Lambda^2(s+2\,b+2\,c)-\dfrac{\pi^2}{4}\Lambda^3(s\,b
+ s\,c +2\,b\,c)\end{split}\eeq{EDR} For Neumann boundary conditions
we obtain \beq
\begin{split}E_{N_{piston}}(\Lambda)&= E_{N_1}(\Lambda)+
E_{N_2}(\Lambda)
\\&=\dfrac{\pi}{8}\Lambda^2(a+b+c)+\dfrac{\pi^2}{4}\Lambda^3(a\,b + a\,c
+b\,c)\\&+\dfrac{\pi}{8}\Lambda^2\big(s-a+b+c\big)+\dfrac{\pi^2}{4}\Lambda^3\big((s-a)\,b
+ (s-a)\,c
+b\,c\big)\\&=\dfrac{\pi}{8}\Lambda^2(s+2\,b+2\,c)+\dfrac{\pi^2}{4}\Lambda^3(s\,b
+ s\,c +2\,b\,c)\,.\end{split}\eeq{ENR} Both
$E_{D_{piston}}(\Lambda)$ and $E_{N_{piston}}(\Lambda)$ have no
dependence on the plate separation $a$. This is due to a
cancellation that has occurred between region I and II. The Casimir
force on the piston has therefore no dependence on the cut-off
$\Lambda$ (since the partial derivative with respect to $a$ of
$E_{D_{piston}}(\Lambda)$ and $E_{N_{piston}}(\Lambda)$ is zero).

\subsection{Cancellation in $d$ dimensions}

In a $d$-dimensional Casimir piston, the piston has $d\!-\!1$
dimensions and divides again the volume into two regions: an
interior region I and exterior region II. Without loss of
generality, the direction in which the piston moves is chosen to be
along the $L_1$ direction so that region I and II share the same
lengths except for $L_1$. It is therefore convenient to write the
Dirichlet and Neumann cut-off expressions \reff{EDCutoff1} and
\reff{ENCutoff1} as a sum of two terms: one that includes $L_1$ and
another which is independent of $L_1$ i.e.  \beq
\begin{split} &E_D(\Lambda)= \dfrac{1}{2^{d+1}}\sum_{m=1}^d (-1
)^{d+m}\,m\,2^m\,\pi^{\frac{m+1}{2}}\,\Gamma(\tfrac{m+1}{2})\,\,\Lambda^{m+1}\,\Big(L_1\,\,\xi^{\,d}_{\,1,\,k_2,..,
k_m} \prod_{i=2}^m\, L_{k_i} +\xi^{\,d}_{\,\substack{k_1,..,
k_m\\k_1\ne1}} \prod_{i=1}^m\,
L_{k_i}\Big)\\&=\dfrac{1}{2^{d+1}}\sum_{m=1}^d (-1
)^{d+m}\,f(m)\,\Big(L_1\,\,\xi^{\,d}_{\,1,\,k_2,.., k_m}
\prod_{i=2}^m\, L_{k_i} +\xi^{\,d}_{\,\substack{k_1,..,
k_m\\k_1\ne1}} \prod_{i=1}^m\, L_{k_i}\Big)
\end{split}\eeq{EDcut_d}
where $f(m)\equiv
\,m\,2^m\,\pi^{\frac{m+1}{2}}\,\Gamma(\tfrac{m+1}{2})\,\,\Lambda^{m+1}$.
The Neumann cut-off expression is given by
 \beq \begin{split} &E_N(\Lambda)\equiv
\dfrac{1}{2^{d+1}}\sum_{m=1}^d
f(m)\,\Big(L_1\,\,\xi^{\,d}_{\,1,\,k_2,.., k_m} \prod_{i=2}^m\,
L_{k_i} +\xi^{\,d}_{\,\substack{k_1,.., k_m\\k_1\ne1}}
\prod_{i=1}^m\, L_{k_i}\Big)\,.
\end{split}\eeq{ENcut_d}
Let the plate separation be $a$. In region I, $L_1=a$ and in region
II, $L_1=s-a$ (the piston splits the length $s$ into $a$ and $s-a$
along the $L_1$ direction) (see fig. 1). To obtain the cut-off
dependence for the $d$-dimensional Casimir piston we need to add the
contributions from regions I and II: \beq
\begin{split} E_{D_{piston}}(\Lambda)& \equiv  E_{D_1}(\Lambda)+E_{D_2}(\Lambda)
\\&=\dfrac{1}{2^{d+1}}\sum_{m=1}^d (-1
)^{d+m}\,f(m)\,\Big(a\,\,\xi^{\,d}_{\,1,\,k_2,.., k_m}
\prod_{i=2}^m\, L_{k_i} +\xi^{\,d}_{\,\substack{k_1,..,
k_m\\k_1\ne1}} \prod_{i=1}^m\, L_{k_i}\Big)
\\&+\dfrac{1}{2^{d+1}}\sum_{m=1}^d (-1
)^{d+m}\,f(m)\,\Big((s-a)\,\,\xi^{\,d}_{\,1,\,k_2,.., k_m}
\prod_{i=2}^m\, L_{k_i} +\xi^{\,d}_{\,\substack{k_1,..,
k_m\\k_1\ne1}} \prod_{i=1}^m\,
L_{k_i}\Big)\\&=\dfrac{1}{2^{d+1}}\sum_{m=1}^d (-1
)^{d+m}\,f(m)\,\Big(s\,\,\xi^{\,d}_{\,1,\,k_2,.., k_m}
\prod_{i=2}^m\, L_{k_i} + 2\,\xi^{\,d}_{\,\substack{k_1,..,
k_m\\k_1\ne1}} \prod_{i=1}^m\, L_{k_i}\Big)
\end{split}\eeq{EDcut_d2}
\beq
\begin{split} E_{N_{piston}}(\Lambda)& \equiv E_{N_1}(\Lambda)+E_{N_2}(\Lambda)
\\&=\dfrac{1}{2^{d+1}}\sum_{m=1}^d f(m) \Big(a\,\,\xi^{\,d}_{\,1,\,k_2,..,
k_m} \prod_{i=2}^m\, L_{k_i} +\xi^{\,d}_{\,\substack{k_1,..,
k_m\\k_1\ne1}} \prod_{i=1}^m\, L_{k_i}\Big)
\\&+\dfrac{1}{2^{d+1}}\sum_{m=1}^d f(m) \Big((s-a)\,\,\xi^{\,d}_{\,1,\,k_2,..,
k_m} \prod_{i=2}^m\, L_{k_i} +\xi^{\,d}_{\,\substack{k_1,..,
k_m\\k_1\ne1}} \prod_{i=1}^m\,
L_{k_i}\Big)\\&=\dfrac{1}{2^{d+1}}\sum_{m=1}^d
\,f(m)\,\Big(s\,\,\xi^{\,d}_{\,1,\,k_2,.., k_m} \prod_{i=2}^m\,
L_{k_i} + 2\,\xi^{\,d}_{\,\substack{k_1,.., k_m\\k_1\ne1}}
\prod_{i=1}^m\, L_{k_i}\Big)
\end{split}\eeq{ENcut_d2}

The cut-off expressions for the piston, $E_{D_{piston}}(\Lambda)$
and $E_{N_{piston}}(\Lambda)$, have no dependence on the plate
separation $a$. Their derivatives with respect to $a$ is zero which
implies that the Casimir force on the piston has no cut-off
dependence in any dimension $d$. The hypersurface divergences have
cancelled out in all dimensions in the Casimir piston scenario.

\section{Casimir force formulas in the $d$-dimensional Casimir
piston}

Having proved the cancellation of hypersurface divergences in the
$d$-dimensional Casimir piston, we now focus on the finite
($\Lambda$-independent) part of the Casimir energy. The finite part
is conveniently expressed as a sum of two terms: an analytical term
composed of a finite sum over Riemann zeta and gamma functions and a
remainder term $R_j$ composed of infinite sums over modified Bessel
functions (though convergence is reached after summing a few terms).
In appendix A we derive exact expressions for the finite part
$E_{0_N}$ and $E_{0_D}$ of the Casimir energy in $d$ dimensions for
Neumann and Dirichlet boundary conditions respectively. In this
section, we state these expressions and use them to obtain the
Neumann and Dirichlet Casimir force $F_N$ and $F_D$ for the
$d$-dimensional Casimir piston. In appendix B we develop alternative
expressions for the Casimir force (see discussion at the end of this
section).

The finite part of the $d$-dimensional Casimir energy for Neumann
and Dirichlet boundary conditions, $E_{0_N}$ and $E_{0_D}$, is given
by \reff{E0N} and \reff{E0D} respectively: \beq
E_{0_N}=-\dfrac{\pi}{2^{d+1}}\!\sum_{m=1}^{d}\!\sum_{j=0}^{m-1}
2^{d-m}\,\xi^{\,m-1}_{\,k_1,.., k_j}\,\dfrac{L_{k_1}\ldots
L_{k_j}}{(L _m)^{j+1}}\Big(
\Gamma(\tfrac{j+2}{2})\,\pi^{\frac{-j-4}{2}}\, \zeta(j+2) +
R_{N_j}\Big) \eeq{E0NN} where the remainder $R_{N_j}$ is given by
\reff{RNj} \beq R_{N_j}
=\sum_{n=1}^{\infty}\,\sumprime_{\substack{\ell_i=-\infty\\i=1,\ldots,
j}}^{\infty}\dfrac{2\,\,n^{\frac{j+1}{2}}}{\pi}\,\dfrac{\,K_{\frac{j+1}{2}}
\big(\,2\pi\,n\,\sqrt{(\ell_1\frac{L_{k_1}}{L_m})^2+\cdots+(\ell_j\,\frac{L_{k_j}}{L_m})^2}\,\,\,\big)}
{\left[(\ell_1\frac{L_{k_1}}{L_m})^2+\cdots+(\ell_j\frac{L_{k_j}}{L_m})^2\right]^{\tfrac{j+1}{4}}}\,,
\eeq{RNjj} and
 \beq E_{0_D}=\dfrac{\pi}{2^{d+1}}\sum_{j=0}^{d-1} (-1
)^{d+j}\,\,\xi^{\,d-1}_{\,k_1,.., k_j}\,\dfrac{L_{k_1}\ldots
L_{k_j}}{(L _d)^{j+1}}\Big(
\Gamma(\tfrac{j+2}{2})\,\pi^{\frac{-j-4}{2}}\, \zeta(j+2) +
R_{D_j}\Big)\,\eeq{E0DD} where $R_{D_j}$ is given by \reff{RDj}:
\beq R_{D_j}
=\sum_{n=1}^{\infty}\,\sumprime_{\substack{\ell_i=-\infty\\i=1,\ldots,
j}}^{\infty}\dfrac{2\,\,n^{\frac{j+1}{2}}}{\pi}\,\dfrac{\,K_{\frac{j+1}{2}}
\big(\,2\pi\,n\,\sqrt{(\ell_1\frac{L_{k_1}}{L_d})^2+\cdots+(\ell_j\,\frac{L_{k_j}}{L_d})^2}\,\,\,\big)}
{\left[(\ell_1\frac{L_{k_1}}{L_d})^2+\cdots+(\ell_j\frac{L_{k_j}}{L_d})^2\right]^{\tfrac{j+1}{4}}}\,.
\eeq{RDjj} The prime on the sum in \reff{RNjj} and \reff{RDjj} means
that the case when all $\ell$'s are simultaneously zero
($\ell_1=\ell_2=\ldots=\ell_j=0$) is to be excluded. There is an
implicit summation over the $k_i$'s via the ordered symbol
$\xi_{\,k_1,.., k_j}$ defined in \reff{ordered}. $R_{N_j}$ and
$R_{D_j}$ do not depend only on $j$ but are also a function of the
ratios of lengths, for example
$R_{N_j}=R_{N_j}(L_{k_1}/L_m,\ldots,L_{k_j}/L_m)$. Therefore, the
implicit summation over the $k_i$'s applies also to $R_{N_j}$ and
$R_{D_j}$. For $j=0$, $R_{N_j}$ and $R_{D_j}$ are defined to be zero
and $\xi_{\,k_1,.., k_j}$ and $L_{k_j}$ are defined to be unity so
that $\xi^{\,d-1}_{\,k_1,.., k_j}\,(L_{k_1}\ldots L_{k_j})/(L
_d)^{j+1}=1/L_d$ for $j=0$.

To obtain the $d$-dimensional Casimir energy in the piston scenario
we need to sum the contributions from region I and region II. In
region I, let the length of the sides of the $d$-dimensional
parallelepiped region be $a_1,a_2,\ldots,a_{d-1},a$ where $a$ is the
plate separation. In region I, we label the lengths $L_i$ such that
$L_1=a_1$, $L_2=a_2$, etc. with $L_d=a$ ($L_d$ is equal to the plate
separation). In region II, the length of the sides are the same as
in region I except for the length $a$ which is replaced by the
length $s-a$. The $d$ lengths are $s-a,a_1,a_2,\ldots,a_{d-1}$. For
region II, we choose to label the lengths $L_i$ such that $L_1=s-a$,
$L_2=a_1$, $L_3=a_2,\ldots, L_d=a_{d-1}$. To calculate the Casimir
force, we only need to keep terms in the Casimir energy that depend
on the plate separation $a$: in region I, this means keeping terms
with $L_d=a$ and in region II this means keeping terms with
$L_1=s-a$.

In region I, the $a$-dependent Casimir energy for Neumann boundary
conditions is obtained by setting $m=d$ so that $L_m=L_d=a$ and
setting $L_{k_j}=a_{k_j}$ in \reff{E0NN}: \beq\begin{split}
E_{0_{N_I}}(a)&=-\dfrac{\pi}{2^{d+1}}\sum_{j=0}^{d-1}
\xi^{\,d-1}_{\,k_1,.., k_j}\,\dfrac{a_{k_1}\ldots
a_{k_j}}{a^{j+1}}\Big(
\Gamma(\tfrac{j+2}{2})\,\pi^{\frac{-j-4}{2}}\, \zeta(j+2) +
R_{I_{N_{j}}}\Big)\end{split} \eeq{E0N1} with $R_{I_{N_j}}$ given by
\beq R_{I_{N_j}}
=\sum_{n=1}^{\infty}\,\sumprime_{\substack{\ell_i=-\infty\\i=1,\ldots,
j}}^{\infty}\dfrac{2\,\,n^{\frac{j+1}{2}}}{\pi}\,\dfrac{\,K_{\frac{j+1}{2}}
\big(\,2\pi\,n\,\sqrt{(\ell_1\frac{a_{k_1}}{a})^2+\cdots+(\ell_j\,\frac{a_{k_j}}{a})^2}\,\,\,\big)}
{\left[(\ell_1\frac{a_{k_1}}{a})^2+\cdots+(\ell_j\frac{a_{k_j}}{a})^2\right]^{\tfrac{j+1}{4}}}\,.
\eeq{RNj1} A word on notation: the Roman numerals I and II will
denote region I and II respectively while $j$ will be denoted via
Arabic numerals $1,2,3$ etc. e.g. $R_{I_{N_1}}$ means the
remainder(R) for Neumann(N) in region I with $j=1$.

The finite part of the Casimir energy for Dirichlet boundary
conditions in region I is obtained by setting $L_d=a$ and
$L_{k_j}=a_{k_j}$ in \reff{E0DD}: \beq
E_{0_{D_I}}=\dfrac{\pi}{2^{d+1}}\sum_{j=0}^{d-1} (-1
)^{d+j}\,\,\xi^{\,d-1}_{\,k_1,.., k_j}\,\dfrac{a_{k_1}\ldots
a_{k_j}}{a^{j+1}}\Big(
\Gamma(\tfrac{j+2}{2})\,\pi^{\frac{-j-4}{2}}\, \zeta(j+2) +
R_{I_{D_j}}\Big)\eeq{E0D1} where  $R_{I_{D_j}}$ is given by
\reff{RDjj}: \beq R_{I_{D_j}}
=\sum_{n=1}^{\infty}\,\sumprime_{\substack{\ell_i=-\infty\\i=1,\ldots,
j}}^{\infty}\dfrac{2\,\,n^{\frac{j+1}{2}}}{\pi}\,\dfrac{\,K_{\frac{j+1}{2}}
\big(\,2\pi\,n\,\sqrt{(\ell_1\frac{a_{k_1}}{a})^2+\cdots+(\ell_j\,\frac{a_{k_j}}{a})^2}\,\,\,\big)}
{\left[(\ell_1\frac{a_{k_1}}{a})^2+\cdots+(\ell_j\frac{a_{k_j}}{a})^2\right]^{\tfrac{j+1}{4}}}\,.
\eeq{RDj1}

In region II, only terms with $L_1=s-a$ contribute to the
$a$-dependent Casimir energy.  We therefore consider only the cases
when $k_1$ is equal to $1$ so that $L_{k_1}=s-a$. The rest of the
lengths ($j>1$) are given by $L_{k_j}=a_{{k_j-1}}$ so that $L_2=a_1,
L_3=a_2, \ldots, L_m=a_{m-1}$. We are interested in an exterior of
infinite length so that the Casimir force in region II is calculated
in the limit $s\!\to\!\infty$. For Neumann boundary conditions, the
case ($j=0, m=1$) in \reff{E0NN} can be omitted because
$L_m=L_1=s-a$ appears in the denominator and yields a zero Casimir
force in the limit of $s\to\infty$. The cases $j=0$, $m>1$ can also
be dropped because they do not yield any terms with $L_1$ (with
$j=0$, the numerator is equal to unity and with $m>1$, $L_m\ne
L_1$). In region II, the lower limit of the sums in \reff{E0NN}
therefore start at $j=1$ and $m=2$ yielding the following Neumann
energy ($a$-dependent part) : \beq
E_{0_{N_{\!I\!I}}}(a)=-\dfrac{\pi}{2^{d+1}}\!\sum_{m=2}^{d}\!\sum_{j=1}^{m-1}
2^{d-m}\,\xi^{\,m-1}_{\,1,k_2,..,
k_j}\,\dfrac{(s-a)\,a_{_{k_2-1}}\ldots a_{_{k_j-1}}}{(a
_{m-1})^{j+1}}\Big( \Gamma(\tfrac{j+2}{2})\,\pi^{\frac{-j-4}{2}}\,
\zeta(j+2) + R_{{I\!I}_{N_j}}\Big) \eeq{E0N2} with \beq
R_{{I\!I}_{N_j}}
=\sum_{n=1}^{\infty}\,\sumprime_{\substack{\ell_i=-\infty\\i=1,\ldots,
j}}^{\infty}\dfrac{2\,\,n^{\frac{j+1}{2}}}{\pi}\,\dfrac{\,K_{\frac{j+1}{2}}
\big(\,2\pi\,n\,\sqrt{(\ell_1\frac{s-a}{a_{m-1}})^2+\cdots+(\ell_j\,\frac{a_{k_j
-1}}{a_{m-1}})^2}\,\,\,\big)}
{\left[(\ell_1\frac{s-a}{a_{m-1}})^2+\cdots+(\ell_j\frac{a_{k_j-1}}{a_{m-1}})^2\right]^{\tfrac{j+1}{4}}}\,.
\eeq{RNj2} For Dirichlet boundary conditions, the case $j=0$ can be
dropped from \reff{E0DD} because it does not yield any terms with
$L_1$. The sum in \reff{E0DD} therefore starts at $j=1$ and we set
$L_{k_1}=s-a$ and $L_{k_j}=a_{_{k_j-1}}$, with $L_d=a_{d-1}$: \beq
E_{0_{D_{\!I\!I}}}(a)=\dfrac{\pi}{2^{d+1}}\sum_{j=1}^{d-1} (-1
)^{d+j}\,\,\xi^{\,d-1}_{\,1,.., k_j}\,\dfrac{(s-a)\ldots
a_{k_j-1}}{(a_{d-1})^{j+1}}\Big(
\Gamma(\tfrac{j+2}{2})\,\pi^{\frac{-j-4}{2}}\, \zeta(j+2) +
R_{{I\!I}_{\!D_j}}\Big)\eeq{E0D2} with \beq R_{{I\!I}_{\!D_j}}
=\sum_{n=1}^{\infty}\,\sumprime_{\substack{\ell_i=-\infty\\i=1,\ldots,
j}}^{\infty}\dfrac{2\,\,n^{\frac{j+1}{2}}}{\pi}\,\dfrac{\,K_{\frac{j+1}{2}}
\big(\,2\pi\,n\,\sqrt{(\ell_1\frac{s-a}{a_{d-1}})^2+\cdots+(\ell_j\,\frac{a_{k_j-1}}{a_{d-1}})^2}\,\,\,\big)}
{\left[(\ell_1\frac{s-a}{a_{d-1}})^2+\cdots+(\ell_j\frac{a_{k_j-1}}{a_{d-1}})^2\right]^{\tfrac{j+1}{4}}}\,.
\eeq{RDj2}

It is now straightforward to calculate the Casimir forces in each
region. The Casimir force contribution from region I for Neumann is
\beq
F_{N_I}=-\dfrac{\partial\,E}{\partial\,a}{}^{{\!0_{N_I}}\!(a)}=-\dfrac{\pi}{2^{d+1}}\sum_{j=0}^{d-1}
\xi^{\,d-1}_{\,k_1,.., k_j}\,(a_{k_1}\ldots
a_{k_j})\,\dfrac{j+1}{a^{j+2}}\,
\Gamma(\tfrac{j+2}{2})\,\pi^{\frac{-j-4}{2}}\,
\zeta(j+2)-\dfrac{\partial R_{I_N}}{\partial \,a}\eeq{FN1} where
$R_{I_N}$ is the remainder contribution given by \beq R_{I_N}=
-\dfrac{\pi}{2^{d+1}}\sum_{j=1}^{d-1} \xi^{\,d-1}_{\,k_1,..,
k_j}\,\dfrac{(a_{k_1}\ldots a_{k_j})}{a^{j+1}} \,\,R_{I_{N_j}}\,.
\eeq{R1N} The corresponding formula for Dirichlet are \beq
F_{D_I}\!=\!-\dfrac{\partial
E}{\partial\,a}{}^{\!0_{D_I}}\!=\!\dfrac{\pi}{2^{d+1}}\sum_{j=0}^{d-1}
(-1 )^{d+j}\,\xi^{\,d-1}_{\,k_1,.., k_j}\,(a_{k_1}\ldots
a_{k_j})\,\dfrac{j+1}{a^{j+2}}\,
\Gamma(\tfrac{j+2}{2})\,\pi^{\frac{-j-4}{2}}
\zeta(j+2)-\dfrac{\partial R_{I_D}}{\partial a}\eeq{FD1} with
$R_{I_D}$ given by \beq R_{I_D}=
\dfrac{\pi}{2^{d+1}}\sum_{j=1}^{d-1}
(-1)^{d+j}\,\xi^{\,d-1}_{\,k_1,.., k_j}\,\dfrac{(a_{k_1}\ldots
a_{k_j})}{a^{j+1}} \,R_{I_{\!D_j}}\,. \eeq{R1D}

The Casimir force from the exterior region II is obtained in the
limit when $s$ tends to infinity. For the Neumann case one obtains
\beq\begin{split}
F_{N_{I\!I}}=-\lim_{s\to\infty}\dfrac{\partial}{\partial\,a}E_{0_{N_{\!I\!I}}}(a)&=-\dfrac{\pi}{2^{d+1}}\!\sum_{m=2}^{d}\!\sum_{j=1}^{m-1}
2^{d-m}\,\xi^{\,m-1}_{\,1,k_2,.., k_j}\,\dfrac{\,a_{_{k_2-1}}\ldots
a_{_{k_j-1}}}{(a _{m-1})^{j+1}}\,
\Gamma(\tfrac{j+2}{2})\,\pi^{\frac{-j-4}{2}}\, \zeta(j+2)
\\&-\dfrac{\pi}{2^{d+1}}\!\sum_{m=3}^{d}\!\sum_{j=2}^{m-1}
2^{d-m}\,\xi^{\,m-1}_{\,1,k_2,.., k_j}\,\dfrac{a_{_{k_2-1}}\ldots
a_{_{k_j-1}}}{(a _{m-1})^{j+1}}\, R_{{I\!I}_{N_j}}\!(\ell_1\!=\!0)
\end{split}\eeq{FN2}
where we used the result\footnote{$-\lim_{s\to\infty}
\,\dfrac{\partial}{\partial\,a} \big[(s-a)\,R_{{I\!I}_{\!N_j}}\big]=
\lim_{s\to\infty}\,R_{{I\!I}_{\!N_j}}-\lim_{s\to\infty}\,(s-a)
\,\dfrac{\partial}{\partial\,a}\,R_{{I\!I}_{N_j}}$. The first term
yields $R_{{I\!I}_{\!N_j}}\!(\ell_1\!=\!0)$ since the modified
Bessel functions decrease to zero exponentially in the limit
$s\!\to\!\infty$ except when $\ell_1\eq0$ (as there is no $s$
dependence when $\ell_1\eq0$). The second term is zero because the
derivative of the modified Bessel functions with respect to $a$
decrease exponentially to zero in the limit $s\!\to\!\infty$ when
$\ell_1\ne 0$. When $\ell_1\eq0$, $R_{{I\!I}_{\!N_j}}$ no longer has
any dependence on $a$ so that its derivative is zero identically.}
$-\lim_{s\to\infty} \,\dfrac{\partial}{\partial\,a}
\big[(s-a)\,R_{{I\!I}_{\!N_j}}\big]\!=\!
R_{{I\!I}_{\!N_j}}\!(\ell_1\eq0)$.
$R_{{I\!I}_{\!N_j}}\!(\ell_1\eq0)$ means  $R_{{I\!I}_{\!N_j}}$
evaluated with $\ell_1\eq0$. Note that the product
$a_{_{k_2-1}}\ldots a_{_{k_j-1}}$ that appears in \reff{FN2} is
identically equal to one for $j\eq1$ so that
$\xi^{\,m-1}_{\,1,k_2,.., k_j}$ $(a_{_{k_2-1}}\ldots
a_{_{k_j-1}})/(a _{m-1})^{j+1}=1/(a _{m-1})^2$ for $j\eq1$.

For the Dirichlet case one obtains
  \beq\begin{split}
F_{D_{I\!I}}=-\lim_{s\to\infty}\dfrac{\partial}{\partial\,a}E_{0_{D_{\!I\!I}}}&=
\dfrac{\pi}{2^{d+1}}\!\sum_{j=1}^{d-1}
(-1)^{d+j}\,\xi^{\,d-1}_{\,1,k_2,..,
k_j}\,\dfrac{\,a_{_{k_2-1}}\ldots a_{_{k_j-1}}}{(a _{d-1})^{j+1}}\,
\Gamma(\tfrac{j+2}{2})\,\pi^{\frac{-j-4}{2}}\, \zeta(j+2)
\\&+\dfrac{\pi}{2^{d+1}}\,\sum_{j=2}^{d-1}
(-1)^{d+j}\,\xi^{\,d-1}_{\,1,k_2,.., k_j}\,\dfrac{a_{_{k_2-1}}\ldots
a_{_{k_j-1}}}{(a _{d-1})^{j+1}}\,
R_{{I\!I}_{\!D_j}}\!(\ell_1\!=\!0)\,.
\end{split}\eeq{FD2}

 The Casimir force
$F_N$ and $F_D$ on the piston for Neumann and Dirichlet respectively
is finally obtained by adding contributions from both region I and
II: \beq F_N=F_{N_I}+F_{N_{I\!I}}\quad;\quad
F_D=F_{D_I}+F_{D_{I\!I}}\,.\eeq{FND} Eq. \reff{FND} together with
\reff{FN1} and \reff{FN2} for  $F_{N_I}$ and  $F_{N_{I\!I}}$, and
\reff{FD1} and \reff{FD2} for $F_{D_I}$ and $F_{D_{I\!I}}$
respectively constitute our final result for the Casimir force on
the piston in $d$ dimensions.

The modified Bessel functions $K_{\frac{j+1}{2}}$ that appear in
$R_{I_{N_{j}}}$ or $R_{I_{D_{j}}}$ (Eqs. \reff{RNj1} and
\reff{RDj1}) converge exponentially fast if the plate separation $a$
is the smallest of the $d$ lengths because the ratios $a_{k_i}/a$
that appear in the argument of the Bessel functions are then greater
than or equal to $1$. Only a few terms need to be summed to reach
high accuracy and the result is also small in magnitude. This is why
we can call $R$ a remainder. However, $R_{I_{N_{j}}}$ and
$R_{I_{D_{j}}}$ can converge slowly and be large if $a$ is larger
than the other lengths. In particular, the large $a$ limit when
$a_{k_i}/a\!<\!<1$ would require a very large number of terms to be
summed before convergence is reached. By making use of the
invariance of the vacuum energy under permutation of lengths, we
derive in appendix B alternative expressions that are more
convenient to use when the plate separation $a$ is large. For
Neumann and Dirichlet they are given by \reff{FaltN22} and
\reff{FaltD22} respectively:
 \beq
F^{\,alt}_N = -\dfrac{\pi}{24\,a^2} -\dfrac{\partial}{\partial a}
R^{\,alt}_{I_N}(\ell_1\!\ne\!0) \eeq{FaltN1} and
 \beq
F^{\,alt}_D = -\dfrac{\partial }{\partial
a}R^{\,alt}_{I_D}(\ell_1\!\ne\!0) \eeq{FaltD1} where
$R^{\,alt}_{I_N}(\ell_1\!\ne\!0)$ is given by \reff{RINAlt} with
\reff{RNj2B2} and $R^{\,alt}_{I_D}(\ell_1\!\ne\!0)$ is given by
\reff{RIDAlt} with \reff{RDj2B2}. The above compact formulas are
applied in the next section in two and three dimensions where one
can see explicitly how they are used.

The ratio of lengths that appear in the argument of the modified
Bessel functions in \reff{RNj2B2} and \reff{RDj2B2} have $a$ in the
numerator ($a/a_{k_i}$) in contrast to our original expressions
(with ratio $a_{k_i}/a$). We have a useful $a \to 1/a$ duality: when
$a$ is large a long computation with the original expressions can be
trivial using the alternative expressions and vice versa when $a$ is
small. The invariance of the vacuum energy under permutations of the
$d$ lengths was used to derive the alternative expressions and the
duality can be traced to this symmetry. Note that regardless of the
size of the plate separation $a$, we would want to label the other
$d\!-\!1$ lengths such that $a_1\ge a_2\ge a_3\ge\ldots\ge a_{d-1}$
to reach the quickest convergence.

The Casimir force on the piston is negative (attractive) in all
dimensions for both Neumann and Dirichlet boundary conditions and
ranges from $-\infty$ (in the limit $a\!\to\!0$) to $0$ (in the
limit $a\!\to\!\infty$). The limit as $a\!\to\!0$ is easily
determined using the original expressions \reff{FND}. In the limit
$a\!\to\!0$, $F_{N_I}$ and $F_{D_I}$ given by \reff{FN1} and
\reff{FD1} tend to  $-1/a^{d+1}$ ( $\partial R_{I_N}/\partial a$ and
$\partial R_{I_D}/\partial a$ tend to zero). $F_{N_{II}}$ and
$F_{D_{II}}$ have no dependence on $a$ so that $F_N=F_{N_I}+
F_{N_{II}}$ and $F_D=F_{D_I}+F_{D_{II}}$ tend towards $-1/a^{d+1}$
and hence $-\infty$ as $a\!\to\!0$. To determine the limit as
$a\!\to\!\infty$, it is easiest to use the alternative expressions.
As already discussed at the end of appendix B, $F^{\,alt}_N$ and
$F^{\,alt}_D$ given by  \reff{FaltN1} and \reff{FaltD1} tend to zero
in that limit because the modified Bessel functions that appear in
$R^{\,alt}_{I_N}(\ell_1\!\ne\!0)$ (Eqs. \reff{RINAlt} and
\reff{RNj2B2}) and $R^{\,alt}_{I_D}(\ell_1\!\ne\!0)$ (Eqs.
\reff{RIDAlt} and \reff{RDj2B2}) decrease exponentially fast to zero
as $a\!\to\!\infty$ (since $\ell_1\!\ne\!0$, when $a\!\to\!\infty$
the argument of the Bessel functions tend to infinity). The rapid
decrease to zero can be seen in the two and three-dimensional plots
of the next section (fig. 2 and fig. 3).

\section{Application: two and three-dimensional Casimir piston for
Neumann boundary conditions}

Exact results for massless scalar fields in the two and
three-dimensional Casimir piston for Dirichlet boundary conditions
were first obtained in Refs. \cite{Cavalcanti,Ariel2} and recently
exact results for 3D Neumann (as well as Dirichlet) were obtained in
\cite{Hertzberg}. We apply our $d$-dimensional formulas (both
expressions) to the two and three-dimensional Casimir piston with
Neumann boundary conditions. The 2D Neumann results are new and fill
a gap in the literature. For 3D Neumann, our first expression looks
similar in form to the one recently derived in \cite{Hertzberg} and
is in numerical agreement with it providing an independent
confirmation of our results. Our alternative 3D Neumann expression
looks quite different in form from the first expression and yields
the same numerical results but is more useful (converges more
quickly) at large plate separation $a$.

\subsection{Two dimensions}

In $d$ dimensions the lengths of the parallelepiped are
$a,a_1,a_2,\ldots,a_{d-1}$ with $a$ being the plate separation. In
two dimensions the lengths are then $a$ and $a_1$ (we set $a_1=b$ so
that the geometry is an $a \times b$ rectangle). The Casimir force
contribution from region I is obtained by setting $d=2$ in
\reff{FN1}: \beq\begin{split} F_{N_I}&=-\dfrac{\pi}{8}\sum_{j=0}^{1}
\xi^{1}_{\,k_1,.., k_j}\,(a_{k_1}\ldots
a_{k_j})\,\dfrac{j+1}{a^{j+2}}\,
\Gamma(\tfrac{j+2}{2})\,\pi^{\frac{-j-4}{2}}\, \zeta(j+2) -
\dfrac{\partial R_{I_N}}{\partial a}
\\&= -\frac{\pi}{48\,a^2}-\frac{\zeta(3)\,b}{8\,\pi\,a^3}+ \dfrac{1}{2}\dfrac{\partial}{\partial a}\sum_{n=1}^{\infty}\sum_{\ell=1}^{\infty}\,\dfrac{n}{\ell}\,
\,\dfrac{1}{a}\,K_1\big(2\,\pi\,n\,\ell\,b/a\big)
\end{split}\eeq{FN21}
where $R_{I_N}$ is obtained from \reff{R1N}:\beq R_{I_N}=-\dfrac{\pi}{8}\,\dfrac{b}{a^2}\,R_{I_{N_1}}(b/a)\\
=-\dfrac{1}{2}\sum_{n=1}^{\infty}\sum_{\ell=1}^{\infty}\,\dfrac{n}{\ell}\,
\,\dfrac{1}{a}\,K_1\big(2\,\pi\,n\,\ell\,b/a\big)\,.\eeq{RIN2D}
$R_{I_{N_1}}(b/a)$ is obtained from \reff{RNj1} and means
$R_{I_{N_1}}$ is a function of $b/a$.

The Casimir force contribution from region II is obtained by setting
$d=2$ in \reff{FN2}. In the first double sum, there is only the term
$j=1, m=2$ to consider. The second double sum is zero (it is nonzero
only starting at $d=3$). We obtain the simple expression \beq
F_{N_{I\!I}}=-\dfrac{\zeta(3)}{16\,\pi\,b^2}\,.\eeq{FN2D}

By summing $F_{N_I}$ and $F_{N_{I\!I}}$ we obtain the Casimir force
$F_N$ on the piston: \beq
F_N=-\frac{\pi}{48\,a^2}-\frac{\zeta(3)\,b}{8\,\pi\,a^3}-\dfrac{\zeta(3)}{16\,\pi\,b^2}+
\dfrac{1}{2}\dfrac{\partial}{\partial
a}\sum_{n=1}^{\infty}\sum_{\ell=1}^{\infty}\,\dfrac{n}{\ell}\,
\,\dfrac{1}{a}\,K_1\big(2\,\pi\,n\,\ell\,b/a\big)\,.\eeq{FN2D} As an
example, the above expression yields $F_N=-0.1342935575/b^2$ for the
case of a square ($a=b$).

\begin{figure}[lt]
\begin{center}
\includegraphics[scale=0.7]{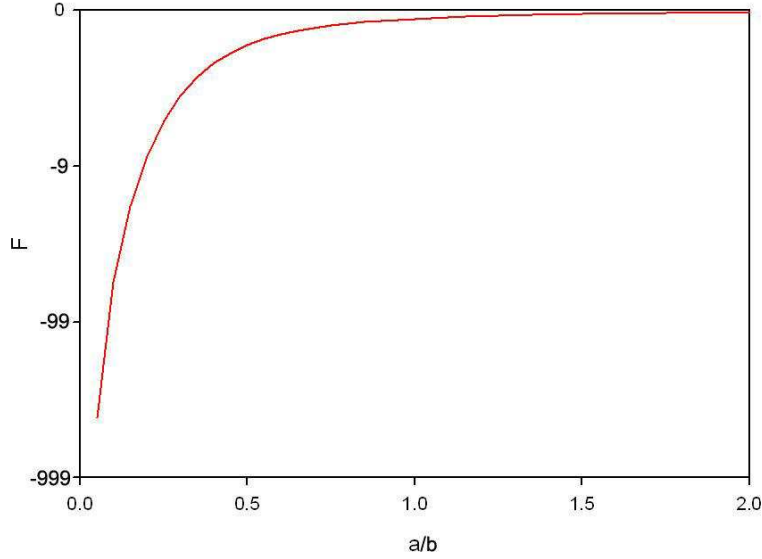}
\caption{Neumann Casimir force $F$ versus $a/b$ where $a$ is the
plate separation and $b$ the second side of a two-dimensional
rectangular region. The force is in units of $1/b^2$. The force is
negative with magnitude decreasing quickly to zero as $a/b$
increases.}
\end{center}
\end{figure}

An alternative expression $F^{\,alt}_N$ for the Casimir force can be
obtained via \reff{FaltN1}. For $d=2$ we obtain \beq
\begin{split}F^{\,alt}_N &= -\dfrac{\pi}{24\,a^2}-\dfrac{\partial
}{\partial a}R^{\,alt}_{I_N}(\ell_1\ne0)
\\&=-\dfrac{\pi}{24\,a^2}+
\dfrac{1}{2\,b}\sum_{n=1}^{\infty}\,\sum_{l=1}^{\infty}\dfrac{n}{\ell}
\,K'_1\big(\,2\pi\,n\,\ell\,a/b\big)\end{split} \eeq{FaltN2D} where
the prime on the modified Bessel function means partial derivative
with respect to $a$ i.e. $K'(x)\equiv\frac{\partial}{\partial a}
K(x)$. $R^{\,alt}_{I_N}(\ell_1\!\ne\!0)$ is obtained from the
$j\!=\!1$, $m\!=\!2$ term in \reff{RINAlt}: \beq
R^{\,alt}_{I_N}(\ell_1\ne0)=-\dfrac{\pi}{8} \dfrac{a}{b^2}\,
R^{\,alt}_{I_{N_1}}(\ell_1\ne0)=-\dfrac{1}{2\,b}\sum_{n=1}^{\infty}\,\sum_{l=1}^{\infty}\dfrac{n}{\ell}
\,K_1\big(\,2\pi\,n\,\ell\,a/b\big)\,.\eeq{RINAltt}
$R^{\,alt}_{I_{N_1}}(\ell_1\!\ne\!0)$ is obtained from the $j\eq1$,
$m\eq2$ term in \reff{RNj2B2}.

The two expressions, \reff{FN2D} and \reff{FaltN2D}, yield the same
value for the Casimir force on the piston and are valid for any
values of $a$ and $b$. However, computationally, expression
\reff{FN2D} is better to use when $a$ is small (i.e. $a/b\!<\!1$),
whereas \reff{FaltN2D} is better to use when $a$ is large
($b/a\!<\!1$). This is the simplest case of the $a\to 1/a$ duality
that was discussed last section.

The Casimir force on the piston is negative (attractive) and ranges
from $-\infty$ (in the limit $a\!\to\!0$) to $0$ (in the limit
$a\!\to\!\infty$). A plot of $F$ versus $a/b$ (in units of $1/b^2$)
is shown in fig. 2.

\subsection{Three dimensions}

In $d$-dimensions we have the $d$ lengths $a,a_1,a_2,\ldots,a_{d-1}$
with $a$ the plate separation. In three dimensions the three lengths
are then $a,a_1$ and $a_2$. For the three-dimensional Casimir piston
it has become customary to use $a,b$ and $c$ for the lengths and we
therefore set $a_2=b$ and $a_1=c$. The Casimir force contribution
from region I is obtained by setting $d=3$ in
\reff{FN1}\beq\begin{split} F_{N_I}&=-\dfrac{\pi}{16}\sum_{j=0}^{2}
\xi^{2}_{\,k_1,.., k_j}\,(a_{k_1}\ldots
a_{k_j})\,\dfrac{j+1}{a^{j+2}}\,
\Gamma(\tfrac{j+2}{2})\,\pi^{\frac{-j-4}{2}}\, \zeta(j+2) -R'_{I_N}
\\&=-\frac{\pi^2\,b\,c}{480\,a^4}-\frac{\zeta(3)\,(b+
c)}{16\,\pi\,a^3} -\frac{\pi}{96\,a^2}-R'_{I_N}
\end{split}\eeq{FN11}
where $R_{I_N}$ is given by \reff{R1N} and \reff{RNj1}: \beq
\begin{split}R_{I_N}&=-\dfrac{\pi}{16}\bigg[\,\dfrac{c}{a^2}\,
R_{I_{N_1}}(c/a)+
\dfrac{b}{a^2}\,R_{I_{N_1}}(b/a)+\dfrac{2\,c}{b^2}\,R_{I_{N_1}}(c/b)\,
+\dfrac{b\,c}{a^3}\,R_{I_{N_2}}(c/a,b/a)\bigg]\\
&\quad=-\dfrac{1}{4}\sum_{n=1}^{\infty}\sum_{\ell=1}^{\infty}\,\dfrac{n}{\ell}\,\Big[
\,\dfrac{1}{a}\,K_1\big(2\,\pi\,n\,\ell\,c/a\big)+ \dfrac{1}{a}\,K_1\big(2\,\pi\,n\,\ell\,b/a\big)+\dfrac{2}{b}\,K_1\big(2\,\pi\,n\,\ell\,c/b\big)\,\Big]\\
&\qquad\qquad-\dfrac{\,b\,c}{8\,a^3}\,\sum_{n=1}^{\infty}\,\sumprime_{\ell_1,\ell_2=-\infty}^{\infty}\!\!
\dfrac{n^{3/2}\,K_{3/2}\Big(2\,\pi\,n\,\sqrt{\Big(\dfrac{\ell_1\,c}{a}\Big)^2+
\Big(\dfrac{\ell_2\,b}{a}\Big)^2}\,\,\Big)}
{\left[\Big(\dfrac{\ell_1\,c}{a}\Big)^2+\Big(\dfrac{\ell_2\,b}{a}\Big)^2\right]^{3/4}}\,.
\end{split}\eeq{RIN}
$R_{I_{N_1}}(c/a)$ means $R_{I_{N_1}}$ is a function of $c/a$ and
the prime above the sum means that the case $\ell_1\eq\ell_2\eq0$ is
to be excluded from the sum.

The Casimir force contribution from region II is obtained by setting
$d=3$ in \reff{FN2}: \beq\begin{split}
F_{N_{I\!I}}&=-\dfrac{\pi}{16}\!\sum_{m=2}^{3}\!\sum_{j=1}^{m-1}
2^{3-m}\,\xi^{\,m-1}_{\,1,k_2,.., k_j}\,\dfrac{\,a_{_{k_2-1}}\ldots
a_{_{k_j-1}}}{(a _{m-1})^{j+1}}\,
\Gamma(\tfrac{j+2}{2})\,\pi^{\frac{-j-4}{2}}\, \zeta(j+2)
-\dfrac{\pi}{16}\,\dfrac{c}{b^3}\,
R_{{I\!I}_{N_2}}\!(\ell_1\!=\!0)\\&= -
\frac{\pi^2\,c}{1440\,b^3}-\frac{\zeta(3)}{16\,\pi}\Big(\frac{1}{c^2}
+\frac{1}{2\,b^2}\Big)
-\dfrac{\,c}{4\,b^3}\,\sum_{n=1}^{\infty}\,\sum_{\ell=1}^{\infty}
\Big(\dfrac{n\,b}{\ell\,c}\Big)^{3/2}K_{3/2}(2\,\pi\,n\,\ell\,c/b)
\end{split}\eeq{FN22}
where \reff{RNj2} with $j\!=\!2$ and $m\!=\!3$ was used to obtain
\beq R_{{I\!I}_{N_2}}\!(\ell_1\!=\!0)=
\dfrac{4}{\pi}\sum_{n=1}^{\infty}\,\sum_{\ell=1}^{\infty}
\Big(\dfrac{n\,b}{\ell\,c}\Big)^{3/2}K_{3/2}(2\,\pi\,n\,\ell\,c/b)\,.
\eeq{r2n2l} The Casimir force $F_N$ on the piston for Neumann
boundary conditions in three dimensions is obtained by summing
$F_{N_I}$ and $F_{N_{I\!I}}$: \beq\begin{split}
F_N&=-\frac{\pi^2\,b\,c}{480\,a^4}-\frac{\zeta(3)\,(b+
c)}{16\,\pi\,a^3} -\frac{\pi}{96\,a^2}-R'_{I_N}\\&-
\frac{\pi^2\,c}{1440\,b^3}-\frac{\zeta(3)}{16\,\pi}\Big(\frac{1}{c^2}
+\frac{1}{2\,b^2}\Big)
-\dfrac{\,c}{4\,b^3}\,\sum_{n=1}^{\infty}\,\sum_{\ell=1}^{\infty}
\Big(\dfrac{n\,b}{\ell\,c}\Big)^{3/2}K_{3/2}(2\,\pi\,n\,\ell\,c/b)
\end{split}\eeq{FND3}
where $R_{I_N}$ is given by \reff{RIN} and $R'_{I_N}\equiv
\partial R_{I_N}/\partial a\,$. Note that the third term in square
brackets in \reff{RIN} depends only on $b$ and $c$ and can be
dropped when evaluating $R'_{I_N}$. For the case of a cube
($a=b=c$), Eq. \reff{FND3} yields $F_N=-0.1380999/c^2$. As in two
dimensions, the force $F_N$ is negative and ranges from $-\infty$
(in the limit $a\!\to\!0$) to $0$ (in the limit $a\!\to\!\infty$). A
3D plot of $F_N$ versus $a/c$ and $b/c$ (in units of $1/c^2$) is
shown in fig. 3.

\begin{figure}[ht]
\begin{center}
\includegraphics[scale=0.8]{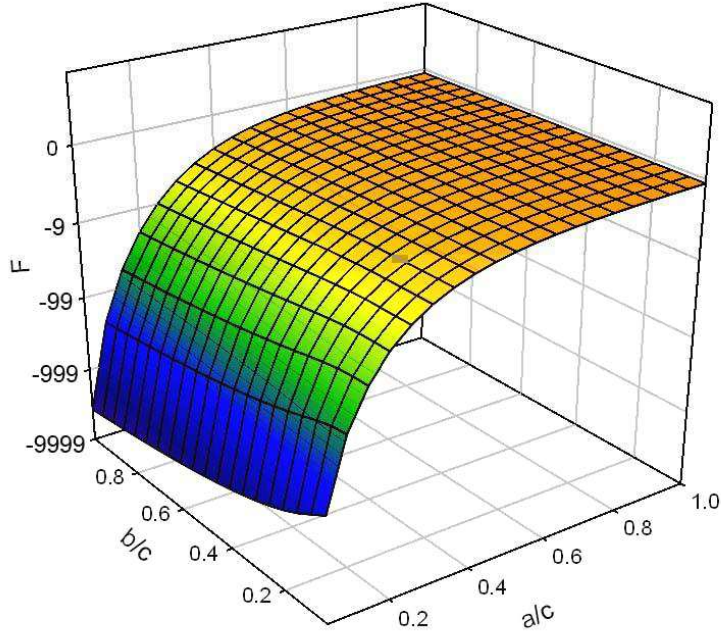}
\caption{3D plot of Neumann Casimir force $F_N$ versus $a/c$ and
$b/c$. The force is in units of $1/c^2$. The force is large and
negative at small values of $a/c$ and remains negative with its
magnitude decreasing quickly to zero as $a/c$ increases. The value
of $b/c$ shifts the magnitude of the force towards larger values as
it increases.}
\end{center}
\end{figure}
Our exact expression \reff{FND3} looks similar in form to the one
derived in \cite{Hertzberg} and is in numerical agreement with it.
Moreover, in the small $a$ limit, $R'_{I_N}$ is exponentially
suppressed (exactly zero in the limit $a\!\to\!0$) and when $b=c$,
the second row in \reff{FND3} yields $0.0429965/c^2$ in agreement
with the Neumann results in both \cite{Kardar} and \cite{Hertzberg}.
This provides an independent confirmation of our results.

An alternative expression $F^{\,alt}_N$ for the Casimir force can be
readily obtained by substituting $d=3$ in \reff{FaltN1} and using
\reff{RINAlt} and \reff{RNj2B2}: \beq
\begin{split}F^{\,alt}_N &= -\dfrac{\pi}{24\,a^2} -\dfrac{\partial
}{\partial a} R^{\,alt}_{I_N}(\ell_1\ne0)\\&= -\dfrac{\pi}{24\,a^2}+
\sum_{n=1}^{\infty}\sum_{\ell=1}^{\infty}\dfrac{n}{4\,\ell}\,
\Big(\dfrac{2}{c}\,K_1^{\prime}(2\,\pi\,n\,\ell\,a/c)+\dfrac{1}{b}\,K_1^{\prime}(2\,\pi\,n\,\ell\,a/b)\Big)
\\&+\dfrac{\partial}{\partial\,a}\Bigg[\,\dfrac{a\,c}{4\,b^3}\sum_{n=1}^{\infty}\sum_
{\ell_1=1}^{\infty}\sum_{\ell_2=-\infty}^{\infty}
n^{3/2}\,\dfrac{K_{3/2}\big(2\,\pi\,n\,\sqrt{(\ell_1\,a/b)^2 +
(\ell_2\,c/b)^2}\big)}{\Big[(\ell_1\,a/b)^2+(\ell_2\,c/b)^2\Big]^{3/4}}\Bigg]\,.
\end{split}\eeq{FaltN11}
 The prime above $K$
denotes partial derivative with respect to $a$. Note that the sum
over $\ell_2$ includes $\ell_2\eq0$ (since the sum over $\ell_2$
contains no prime above it). Equation \reff{FaltN11} is our
alternative expression for the exact Casimir force on the piston in
three dimensions for Neumann boundary conditions. It is valid for
any values of the lengths $a,b$ and $c$ and yields the same Casimir
force as the original expression \reff{FND3}. However, in the
argument of the modified Bessel functions the plate separation $a$
now appears in the numerator making the alternative expression
converge quickly for large $a$. A long computation with the original
expression when $a$ is large can converge exponentially fast with
the alternative expression and vice versa when $a$ is small. This is
a nontrivial case of the $a \to 1/a$ duality already encountered in
two dimensions and discussed in general in the last section. Note
that we are free to label the base such that $c\ge b$ to obtain the
best convergence.

\section{Summary and discussion}

By applying a multidimensional cut-off technique we obtained
expressions for the cut-off dependent part of the Casimir energy for
parallelepiped geometries in any spatial dimension $d$ and showed
explicitly that nonrenormalizable hypersurface divergences cancel in
the Casimir piston scenario in all dimensions. We then obtained
exact expressions for the $d$-dimensional Casimir force on a piston
for the case of massless scalar fields obeying Dirichlet and Neumann
boundary conditions. As an example, we applied the $d$-dimensional
formulas to the 2D and 3D piston with Neumann boundary conditions.
The two main features of the Casimir piston originally mentioned by
Cavalcanti \cite{Cavalcanti} for a 2D rectangular geometry, namely
the cancellation of the surface divergences and the negative Casimir
force on the piston, were shown to hold true in all dimensions $d$.
We obtained two different expressions for the $d$-dimensional
Casimir force. The Casimir energy is clearly invariant under
permutations of the $d$ lengths of the parallelepiped. This symmetry
is trivial but its application is very useful: one can derive
alternative expressions for the Casimir force that converge quickly
compared to the original expressions when the plate separation $a$
is large.

It would be interesting to generalize our $d$-dimensional results to
include arbitrary cross sections and thermal corrections. For scalar
fields in three dimensions this has been recently considered in
\cite{Hertzberg} via the optical path technique. The scalar field
results were then used to obtain the electromagnetic (EM) Casimir
energies with perfect metallic boundary conditions \cite{Hertzberg}.
It would be worthwhile to see how the 3D alternative expressions for
massless scalar fields derived here for  Neumann and for Dirichlet
elsewhere \cite{Ariel2} can be modified to include thermal
corrections. These could then be used to obtain 3D alternative
expressions for the thermal corrections to the EM case.

\begin{appendix}
\section{Cut-off dependent and finite parts of the regularized vacuum energy: periodic, Dirichlet and Neumann boundary conditions}
\def\theequation{A.\arabic{equation}}
\setcounter{equation}{0}

We consider a massless scalar field confined to a $d$-dimensional
parallelepiped region with arbitrary lengths $L_1,...,L_d$ obeying
periodic, Neumann and Dirichlet boundary conditions. Our goal is to
include the cut-off dependent and finite parts of the vacuum energy
regularized via a multidimensional cut-off technique \cite{Ariel}.
This appendix naturally divides into two parts. We first consider
periodic boundary conditions and make use of formulas found in
section 2 and appendix B of \cite{Ariel}. In particular, we
determine explicitly the $d$-dimensional cut-off dependence in the
expression for the regularized vacuum energy. In contrast to
dimensional or zeta-function regularization, the multidimensional
cut-off technique performs no renormalization. The second part
consists of finding the regularized vacuum energy for the Dirichlet
and Neumann cases. This is obtained by summing over the vacuum
energy of the periodic case.

The vacuum energy for periodic boundary conditions regularized using
a cut-off $\lambda$ is given by \cite{Ariel} \beq\begin{split}&
\tilde{E}_p\,(d,\lambda) =\,\pi
\sum_{\substack{n_i=-\infty\\i=1,\ldots, d}}^{\infty}
\sqrt{\tfrac{n_1^2}{L_1^2} \,+\, \cdots\, +\, \tfrac{n_d^2}{L_d^2}}
\,\,\expo{-\lambda\,\sqrt{\tfrac{n_1^2}{L_1^2} \,+\, \cdots\, +\,
\tfrac{n_d^2}{L_d^2}}}= -\pi \,\partial_{\lambda}\!
\sum_{\substack{n_i=-\infty\\i=1,\ldots, d}}^{\infty}
\expo{-\lambda\,\sqrt{\tfrac{n_1^2}{L_1^2} \,+\, \cdots\, +\,
\tfrac{n_d^2}{L_d^2}}}
\\&=-\pi\,\partial_{\lambda}\,\Big(1\! +\!
\sumprime_{n_1=-\infty}^{\infty}\!\!\expo{-\lambda\,\sqrt{\tfrac{n_1^2}{L_1^2}}}
\!+\!\sumprime_{n_2=-\infty}^{\infty}\sum_{n_1=-\infty}^{\infty}\!\!\expo{-\lambda\,\sqrt{\tfrac{n_1^2}{L_1^2}
\,+\, \tfrac{n_2^2}{L_2^2}}} \,+\, \cdots \,+\,
\sumprime_{n_d=-\infty}^{\infty}\sum_{\substack{n_i=-\infty\\i=1,\ldots,
d-1}}^{\infty}\!\!\!\expo{-\lambda\,\sqrt{\tfrac{n_1^2}{L_1^2}\, +\,
\cdots \,+\, \tfrac{n_d^2}{L_d^2}}}\,\Big)\\&= -\pi
\,\sum_{j=0}^{d-1}\partial_{\lambda}\,\Lambda_j(\lambda)
\end{split}\eeq{start2B}
where \beq
\Lambda_j(\lambda)\equiv\sumprime_{n=-\infty}^{\infty}\,\sum_{\substack{n_i=-\infty\\i=1,\ldots,
j}}^{\infty}\!\!\!\expo{-\lambda\,\sqrt{\tfrac{n^2}{L_{j+1}^2}\,+\,\tfrac{n_1^2}{L_1^2}
\,+\, \cdots + \tfrac{n_j^2}{L_j^2}}}\,. \eeq{lambdajB}

The prime on the sum over $n$ means that $n=0$ is excluded from the
sum. The function $\partial_{\lambda}\,\Lambda_j(\lambda)$ can be
expressed in the following form \cite{Ariel} (in the limit $\lambda
\to 0$) \beq\partial_{\lambda}\,\Lambda_j(\lambda)=\dfrac{L_1\ldots
L_j}{(L
_{j+1})^{j+1}}\,\Big(\,2^{\,j+1}\,\partial_{\lambda'}\!\!\sum_{n=1}^{\infty}\int_{0}^{\infty}\,\expo{-\lambda'\,\sqrt{n^2+
x_1^2 \,+\, \cdots + x_j^2}}\,dx_1\ldots dx_j + R_j\Big) \eeq{jsum1}
where $\lambda'\equiv \lambda/L_{j+1}$ and $L_1\ldots L_j$ is a
product of lengths i.e. $\prod_{i=1}^j L_i$. This product is defined
to be unity for the special case of $j=0$. $R_j$ is given by
\cite{Ariel} \beq\begin{split} R_j
&=\sum_{n=1}^{\infty}\,\sumprime_{\substack{\ell_i=-\infty\\i=1,\ldots,
j}}^{\infty}\dfrac{2\,(n\,L_{j+1})^{\tfrac{j+1}{2}}} {\pi\,
\left[(\ell_1\,L_1)^2+\cdots+(\ell_j\,L_j)^2\right]^{\tfrac{j+1}{4}}}\,K_{\tfrac{j+1}{2}}
\big(\,\tfrac{2\pi\,n}{L_{j+1}}\sqrt{(\ell_1\,L_1)^2+\cdots+(\ell_j\,L_j)^2}\,\,\,\big)\,.
\end{split}
\eeq{rjBB} $R_j$ starts at $j=1$ (it is zero for $j=0$). The
functions $K_{(j+1)/2}$ are modified Bessel functions and the prime
on the sum means that the case where all the $\ell_i$'s are zero is
excluded. Via the Euler-Maclaurin formula, the integral term in the
round brackets in \reff{jsum1} can be decomposed into a cut-off
dependent term (which diverges as $\lambda \to 0$) and a finite term
which is independent of the cut-off (see section 2 of \cite{Ariel}):
\beq\begin{split}
&2^{\,j+1}\,\partial_{\lambda'}\!\!\sum_{n=1}^{\infty}\int_{0}^{\infty}\,\expo{-\lambda'\,\sqrt{n^2+
x_1^2 \,+\, \cdots + x_j^2}}\,dx_1\ldots dx_j \\ & =
\Gamma(\tfrac{j+2}{2})\,\pi^{\frac{-j-4}{2}}\,\zeta(j+2)\,\,+\,\,\dfrac{j}{\lambda'{^{j+1}}}\,2^j
\,\pi^{\frac{j-1}{2}}\, \Gamma(\tfrac{j+1}{2})
-\dfrac{j+1}{\lambda'^{j+2}}\,2^{j+1}\,\pi^{\frac{j}{2}}\,\Gamma(\tfrac{j+2}{2})\,
\\& =
\Gamma(\tfrac{j+2}{2})\,\pi^{\frac{-j-4}{2}}\,\zeta(j+2)\,\,+\,\,\dfrac{j}{\lambda^{j+1}}\,2^j
\,\pi^{\frac{j-1}{2}}\, \Gamma(\tfrac{j+1}{2})\,(L_{_{j+1}})^{\,j+1}
-\dfrac{(j+1)}{\lambda^{j+2}}\,2^{j+1}\,\pi^{\frac{j}{2}}\,\Gamma(\tfrac{j+2}{2})\,(L_{_{j+1}})^{\,j+2}\,.
\end{split}\eeq{first} Substituting \reff{first} into \reff{jsum1}, the regularized vacuum energy \reff{start2B} for periodic boundary conditions
is given by ($\word{as} \lambda\to0$)\beq
\begin{split}&\tilde{E}_{p_{ _{L_1\ldots L_d}}}(d,\lambda)
=-\,\pi\sum_{j=0}^{d-1}\dfrac{L_1\ldots L_j}{(L _{j+1})^{j+1}}\Bigg[
\Gamma(\tfrac{j+2}{2})\,\pi^{\frac{-j-4}{2}}\, \zeta(j+2) + R_j
\\&\qquad\qquad\qquad\qquad\qquad\qquad+ \dfrac{j}{\lambda^{j+1}}\,2^j \,\pi^{\frac{j-1}{2}}\,
\Gamma(\tfrac{j+1}{2})\,(L_{_{j+1}})^{\,j+1}
-\dfrac{(j+1)}{\lambda^{j+2}}\,2^{j+1}\,\pi^{\frac{j}{2}}\,\Gamma(\tfrac{j+2}{2})\,(L_{_{j+1}})^{\,j+2}\Bigg]
\\&=-\,\pi\sum_{j=0}^{d-1}\dfrac{L_1\ldots L_j}{(L _{j+1})^{j+1}}\Big(
\Gamma(\tfrac{j+2}{2})\,\pi^{\frac{-j-4}{2}}\, \zeta(j+2) +
R_j\Big)+(L_1\ldots
L_d)\dfrac{d}{\lambda^{d+1}}\,2^d\,\pi^{\frac{d+1}{2}}\,\Gamma(\tfrac{d+1}{2})
\\&=\dfrac{-\pi}{6\,L_1}-\dfrac{\zeta(3)}{2\,\pi}\dfrac{L_1}{L_2^2}-\dfrac{\pi^2}{90}\dfrac{L_1\,L_2}{L_3^3}
\!+\!\cdots
-R_1\,\dfrac{\pi\,L_1}{L_2^2}-R_2\,\dfrac{\pi\,L_1\,L_2}{L_3^3}\!+\!\cdots\!+\!\dfrac{d}{\lambda^{d+1}}\,2^d\,\pi^{\frac{d+1}{2}}\,\Gamma(\tfrac{d+1}{2})(L_1\ldots
L_d)
\end{split}\eeq{epfinal}
The notation $\tilde{E}_{p_{ _{\,L_1,\ldots, L_d}}}\!\!(d,\lambda)$
is a compact way of stating that the regularized vacuum energy
$\tilde{E}_p$ is a function of the dimension $d$, the cut-off
parameter $\lambda$ and the lengths $L_1,...,L_d$.

The regularized vacuum energy for Dirichlet and Neumann boundary
conditions can be expressed as a sum over the periodic energies
$\tilde{E}_p$ \cite{Ariel2, Wolfram}: \beq \tilde{E}_{\binom{N}{D}}
= \dfrac{1}{2^{d+1}}\sum_{m=1}^d \,(\pm \,1 )^{d+m}
\,\,\xi^{\,d}_{\,k_1,..,
k_m}\,\tilde{E}_{p_{_{\,L_{k_1},..,L_{k_m}}}}(m,\lambda)
\eeq{Edirich} where the plus and negative signs correspond to the
Neumann ($N$) and Dirichlet ($D$) cases respectively.
$\xi^{\,d}_{\,k_1,.., k_m}$ is called the ordered symbol and is
defined by \cite{Ariel3} \beq \xi^{\,d}_{\,k_1,..,
k_m}=\begin{cases} 1&\word{if}
 k_1 \less k_2 \less \ldots\! <k_m \,;\,1 \le k_m \le d\\ 0&
\word{otherwise}.
\end{cases}\eeq{final}
The $k_i$'s are positive integers that can run from $1$ to a maximum
value of $d$. The ordered symbol $\xi^{\,d}_{\,k_1,.., k_m}$ ensures
that the $implicit$ $summation$ over the $k_i$'s is over all
distinct sets $\{k_1,.., k_m\}$ under the constraint that
$k_1\!<\!k_2\!<\!\cdots\!<k_m$. The superscript $d$ specifies the
maximum value of $k_m$. $E_{p_{_{L_{k_1},..,L_{k_m}}}}(\lambda,m)$
is obtained from \reff{epfinal} by replacing $d$ by $m$ and $L_1$ by
$L_{k_1}$, $L_2$ by $L_{k_2}$, etc. When substituted into
\reff{Edirich} one obtains \beq \begin{split}E_{\binom{N}{D}} &=
\dfrac{-\pi}{2^{d+1}}\sum_{m=1}^d \,(\pm \,1 )^{d+m}
\xi^{\,d}_{\,k_1,.., k_m} \,\sum_{j=0}^{m-1} \dfrac{L_{k_1}\ldots
L_{k_j}}{(L _{k_{j+1}})^{j+1}}\Big(
\Gamma(\tfrac{j+2}{2})\,\pi^{\frac{-j-4}{2}}\, \zeta(j+2) + R_j\Big)
\\&+\dfrac{1}{2^{d+1}}\sum_{m=1}^d \,(\pm \,1 )^{d+m}
\,\,\xi^{\,d}_{\,k_1,.., k_m}\,(L_{k_1}\ldots
L_{k_m})\dfrac{m}{\lambda^{m+1}}\,2^m\,\pi^{\frac{m+1}{2}}\,\Gamma(\tfrac{m+1}{2})\end{split}\eeq{Edirich3}
where $R_j$ is now the function \reff{rjBB} with $L_1$ replaced by
$L_{k_1}$, $L_2$ by $L_{k_2}$, etc. It is convenient to define the
function  \beq f_{j_{_{\,\,k_1,..,k_{j+1}}}} \equiv
\dfrac{L_{k_1}\ldots L_{k_j}}{(L _{k_{j+1}})^{j+1}}\Big(
\Gamma(\tfrac{j+2}{2})\,\pi^{\frac{-j-4}{2}}\, \zeta(j+2) +
R_j\Big)\,.\eeq{fk1} Using \reff{fk1} and rearranging the limits on
$m$ and $j$, we can express \reff{Edirich3} in the following compact
form (we write out separately the Dirichlet and Neumann cases)
\begin{eqnarray} \tilde{E}_D&=&\dfrac{-\pi}{2^{d+1}}\sum_{j=0}^{d-1}
\sum_{m=j+1}^{d}(-1)^{d+m}\,\xi^{\,d}_{\,k_1,.., k_m}
f_{j_{_{\,\,k_1,..,k_{j+1}}}}
+E_D(\Lambda)\label{EDD1}\\\tilde{E}_N&=&\dfrac{-\pi}{2^{d+1}}\sum_{j=0}^{d-1}
\sum_{m=j+1}^{d}\xi^{\,d}_{\,k_1,.., k_m}
f_{j_{_{\,\,k_1,..,k_{j+1}}}} +E_N(\Lambda)
\label{ENN1}\end{eqnarray} where the functions $E_D(\Lambda)$ and
$E_N(\Lambda)$ are the cut-off dependent terms for the Dirichlet and
Neumann cases respectively obtained from the second row in
\reff{Edirich3} (we now work with the cut-off $\Lambda\equiv
1/\lambda$ instead of $\lambda$ so that the divergent limit
$\lambda\to 0$ is replaced by $\Lambda\to\infty$ which is the more
customary notation):
\begin{eqnarray} E_D(\Lambda)&\equiv&\dfrac{1}{2^{d+1}}\sum_{m=1}^d
\,(-1 )^{d+m} \,\,\xi^{\,d}_{\,k_1,.., k_m}\,(L_{k_1}\ldots
L_{k_m})\,m\,2^m\,\pi^{\frac{m+1}{2}}\,\Gamma(\tfrac{m+1}{2})\,\Lambda^{m+1}
\label{EDCutoff}\\E_N(\Lambda)&\equiv&\dfrac{1}{2^{d+1}}\sum_{m=1}^d
\,\xi^{\,d}_{\,k_1,.., k_m}\,(L_{k_1}\ldots
L_{k_m})\,m\,2^m\,\pi^{\frac{m+1}{2}}\,\Gamma(\tfrac{m+1}{2})\,\Lambda^{m+1}
\label{ENCutoff}\end{eqnarray} Note that in \reff{ENN1} the limits
on $m$ and $j$ in the double sum have been rearranged compared to
\reff{Edirich3}. We can decompose $\xi^{\,d}_{\,k_1,.., k_m}$ into a
sum of two terms: $\xi^{\,d-1}_{\,k_1,..,
k_{m-1},d}+\xi^{\,d-1}_{\,k_1,.., k_m}$. In the first term, $k_m$ is
set to its maximum value of $d$ and the implicit sum is over the
remaining $k_i$'s with the maximum value of $k_{m-1}$ equal to $d-1$
(hence the superscript $d-1$). The second term contains the
remaining implicit summation with the maximum value of $k_m$ equal
to $d-1$ (hence there is a superscript $d\!-\!1$ in the second term
as well). For the case $m=d$, the decomposition yields only one term
$\xi^{\,d}_{\,k_1,.., k_d}=\xi^{\,d-1}_{\,k_1,.., k_{d-1},d} + 0$
since $k_d$ can only be equal to $d$.

With this decomposition the sum over $m$ becomes
\beq\sum_{m=j+1}^{d}(\pm \,1 )^{d+m}\,\, \xi^{\,d}_{\,k_1,.., k_m} =
\sum_{m=j+1}^{d}(\pm \,1 )^{d+m}\,\, \big[\,\xi^{\,d-1}_{\,k_1,..,
k_{m-1},d} + \xi^{\,d-1}_{\,k_1,.., k_{m}}\,\big]\,. \eeq{thatsit}
There are two separate cases to evaluate above: the plus sign (the
Neumann case) and the minus sign (the Dirichlet case). The Dirichlet
case has already been calculated in \cite{Ariel3} and the result is
\beq \sum_{m=j+1}^{d}(-1 )^{d+m}\,\, \xi^{\,d}_{\,k_1,.., k_m}
=(-1)^{d+j+1}\,\xi^{\,d-1}_{\,k_1,.., k_j,d}\,.\eeq{xiD} For the
Neumann case one obtains
\beq\begin{split}\sum_{m=j+1}^{d}\xi^{\,d}_{\,k_1,.., k_m}
&=\xi^{\,d-1}_{\,k_1,.., k_j,d} + (\,\xi^{\,d-1}_{\,k_1,..,
k_{j+1}}+ \xi^{\,d-1}_{\,k_1,..,
k_{j+1},d}\,)\\&\qquad\qquad+(\,\xi^{\,d-1}_{\,k_1,.., k_{j+2}}+
\xi^{\,d-1}_{\,k_1,..,
k_{j+2},d}\,)+\ldots+(\,\xi^{\,d-1}_{\,k_1,.., k_{d-1}} +
\xi^{\,d-1}_{\,k_1,.., k_{d-1},d})\,.\end{split} \eeq{xiDNeum} Each
pair of round brackets contains the sum of two terms which are
equal. For example, consider the first pair of round brackets
$(\xi^{\,d-1}_{\,k_1,.., k_{j+1}}+\xi^{\,d-1}_{\,k_1,..,
k_{j+1},d})\,$. The fact that $k_{j+2}$ is equal to $d$ in the
second term is irrelevant since the summation over
$f_{j_{_{\,\,k_1,..,k_{j+1}}}}$ in \reff{ENN1} ends at $k_{j+1}$ for
a given $j$. Therefore, $\xi^{\,d-1}_{\,k_1,.., k_{j+1},d}$ is equal
to $\xi^{\,d-1}_{\,k_1,.., k_{j+1}}$. The same logic applies to the
other pairs of round brackets. Equation \reff{xiDNeum} reduces to a
recursion relation
\beq\begin{split}\sum_{m=j+1}^{d}\xi^{\,d}_{\,k_1,..,
k_m}&=\xi^{\,d-1}_{\,k_1,.., k_j,d} +
2\!\!\sum_{m=j+1}^{d-1}\xi^{\,d-1}_{\,k_1,.., k_m}
\\&=\xi^{\,d-1}_{\,k_1,.., k_j,d} + 2\,\big(\xi^{\,d-2}_{\,k_1,..,
k_j,d-1} + 2\!\!\sum_{m=j+1}^{d-2}\xi^{\,d-2}_{\,k_1,.., k_m}\big)
\,.\end{split}\eeq{ret} Applying the above recursion repeatedly
(another $d\!-\!j\!-2$ times) yields \beq
\sum_{m=j+1}^{d}\xi^{\,d}_{\,k_1,.., k_m} = \sum_{m=j+1}^{d}
2^{d-m}\,\xi^{\,m-1}_{\,k_1,.., k_j,m}\,.\eeq{xiDNeum2} With
\reff{xiDNeum2}, the double sum in \reff{ENN1} can be expressed
as\beq \begin{split} \dfrac{-\pi}{2^{d+1}}\sum_{j=0}^{d-1}
\sum_{m=j+1}^{d}\,2^{d-m}\,\xi^{\,m-1}_{\,k_1,.., k_j,m}
\,f_{j_{_{\,\,k_1,..,k_{j+1}}}}
\!\!\!=\dfrac{-\pi}{2^{d+1}}\sum_{m=1}^{d}\sum_{j=0}^{m-1}
\,2^{d-m}\,\xi^{\,m-1}_{\,k_1,.., k_j}
\,f_{j_{_{\,\,k_1,..,k_j,\,m}}}\,.\end{split} \eeq{fN}  The function
$f_{j_{_{\,\,k_1,\,..,\,k_j,\,m}}}$ is given by \reff{fk1} with
$k_{j+1}$ equal to $m$. Substituting \reff{fN} into \reff{ENN1}
yields our final expression for the Neumann regularized vacuum
energy \beq \tilde{E}_N\!=E_{0_N}+ E_N(\Lambda)\eeq{Neum-cut-off}
where the finite part $E_{0_N}$ is given by \beq
E_{0_N}=\dfrac{-\pi}{2^{d+1}}\!\sum_{m=1}^{d}\!\sum_{j=0}^{m-1}
2^{d-m}\,\xi^{\,m-1}_{\,k_1,.., k_j}\,\dfrac{L_{k_1}\ldots
L_{k_j}}{(L _m)^{j+1}}\Big(
\Gamma(\tfrac{j+2}{2})\,\pi^{\frac{-j-4}{2}}\, \zeta(j+2) +
R_{N_j}\Big)\,. \eeq{E0N} The function $R_{N_j}$ is given by
\reff{rjBB} with $L_1\to L_{k_1}$, $L_{j+1}\to L_{k_{j+1}}=L_m$:
\beq R_{N_j}
=\sum_{n=1}^{\infty}\,\sumprime_{\substack{\ell_i=-\infty\\i=1,\ldots,
j}}^{\infty}\dfrac{2\,\,n^{\frac{j+1}{2}}}{\pi}\,\dfrac{\,K_{\frac{j+1}{2}}
\big(\,2\pi\,n\,\sqrt{(\ell_1\frac{L_{k_1}}{L_m})^2+\cdots+(\ell_j\,\frac{L_{k_j}}{L_m})^2}\,\,\,\big)}
{\left[(\ell_1\frac{L_{k_1}}{L_m})^2+\cdots+(\ell_j\frac{L_{k_j}}{L_m})^2\right]^{\tfrac{j+1}{4}}}\,.
\eeq{RNj} The Dirichlet case is obtained by substituting \reff{xiD}
in \reff{EDD1} yielding \beq
\tilde{E}_D=\dfrac{\pi}{2^{d+1}}\sum_{j=0}^{d-1} (-1
)^{d+j}\,\xi^{\,d-1}_{\,k_1,.., k_j}\,
f_{j_{_{\,\,k_1,\,..,\,k_j,\,d}}} +E_D(\Lambda)\,. \eeq{fD} The
function $f_{j_{_{\,\,k_1,\,..,\,k_j,\,d}}}$ is obtained by setting
$k_{j+1}$ equal to $d$ in \reff{fk1} yielding the Dirichlet
regularized vacuum energy \beq \tilde{E}_D = E_{0_D}
+E_D(\Lambda)\eeq{ED22} where the finite part $E_{0_D}$ is given by
\beq E_{0_D}=\dfrac{\pi}{2^{d+1}}\sum_{j=0}^{d-1} (-1
)^{d+j}\,\,\xi^{\,d-1}_{\,k_1,.., k_j}\,\dfrac{L_{k_1}\ldots
L_{k_j}}{(L _d)^{j+1}}\Big(
\Gamma(\tfrac{j+2}{2})\,\pi^{\frac{-j-4}{2}}\, \zeta(j+2) +
R_{D_j}\Big)\,.\eeq{E0D} The function $R_{D_j}$ is given by
\reff{rjBB} with $L_1\to L_{k_1}$, $L_{j+1}\to L_{k_{j+1}}=L_d$:
\beq R_{D_j}
=\sum_{n=1}^{\infty}\,\sumprime_{\substack{\ell_i=-\infty\\i=1,\ldots,
j}}^{\infty}\dfrac{2\,\,n^{\frac{j+1}{2}}}{\pi}\,\dfrac{\,K_{\frac{j+1}{2}}
\big(\,2\pi\,n\,\sqrt{(\ell_1\frac{L_{k_1}}{L_d})^2+\cdots+(\ell_j\,\frac{L_{k_j}}{L_d})^2}\,\,\,\big)}
{\left[(\ell_1\frac{L_{k_1}}{L_d})^2+\cdots+(\ell_j\frac{L_{k_j}}{L_d})^2\right]^{\tfrac{j+1}{4}}}\,.
\eeq{RDj} For $j=0$, $R_{N_j}$ and $R_{D_j}$ are defined to be zero
and $\xi_{\,k_1,.., k_j}$ and $L_{k_j}$ are defined to be unity.

The final expressions for the regularized vacuum energy are
\reff{Neum-cut-off} for the Neumann case and \reff{ED22} for the
Dirichlet case with the cut-off dependent parts $E_D(\Lambda)$ and
$E_N(\Lambda)$ given by \reff{EDCutoff} and \reff{ENCutoff} and the
finite parts $E_{0_N}$ and $E_{0_D}$ given by \reff{E0N} and
\reff{E0D} respectively.

\section{Alternative expressions for the $d$-dimensional Casimir piston}
\def\theequation{B.\arabic{equation}}
\setcounter{equation}{0}

In section 3 we derived expressions for the Casimir force on the
piston. In this appendix we derive alternative expressions. This is
accomplished by labeling the lengths $L_i$ in region I differently
compared to section 3. The Casimir energy is invariant under
permutation of lengths so a different labeling does not alter the
value of the Casimir energy. However, the different labeling leads
to an expression with a different form.

In section 3 we labeled the lengths $L_i$ in region I as
$L_1\!=\!a_1, L_2\!=\!a_2,\ldots,L_{d-1}\!=\!a_{d-1}$ with $L_d$
equal to the plate separation $a$. We now label $L_i$ such that
$L_1=a, L_2=a_1, L_3=a_2,\ldots,L_d=a_{d-1}$ so that $L_1$ is equal
to the plate separation.  We do not change the labeling in region II
(i.e. $L_1=s-a,L_2=a_1,L_3=a_2$, etc.) so that we only need to
obtain new formulas for region I.

The expressions for the finite part of the Casimir energy are given
by \reff{E0NN} and \reff{E0DD} for Neumann and Dirichlet
respectively. We only keep the $a$-dependent terms and for region I
this means keeping only those terms with $L_1=a$. For Neumann,
\reff{E0NN} divides into two sums: the $j=0$, $m=1$ term where
$L_m=L_1=a$ appears in the denominator and all other terms where
$L_1$ appears in the numerator (this occurs when $k_1=1$ so that
$L_{k_1}=a$ for $j>0$ with the other lengths given by
$L_{k_j}=a_{k_j-1}$). This yields (with ``alt" as superscript for
``alternative") \beq\begin{split}
E^{\,alt}_{0_{N_I}}(a)&=-\dfrac{\pi}{2^{d+1}}\!\sum_{m=1}^{d}\!\sum_{j=0}^{m-1}
2^{d-m}\,\xi^{\,m-1}_{\,k_1,.., k_j}\,\dfrac{L_{k_1}\ldots
L_{k_j}}{(L _m)^{j+1}}\Big(
\Gamma(\tfrac{j+2}{2})\,\pi^{\frac{-j-4}{2}}\, \zeta(j+2) +
R_{N_j}\Big) \\&=
-\dfrac{\pi}{24\,a}-\dfrac{\pi}{2^{d+1}}\!\sum_{m=2}^{d}\!\sum_{j=1}^{m-1}
2^{d-m}\,\xi^{\,m-1}_{\,1,k_2,.., k_j}\,\dfrac{a\,a_{_{k_2-1}}\ldots
a_{_{k_j-1}}}{(a _{m-1})^{j+1}}\Big(
\Gamma(\tfrac{j+2}{2})\,\pi^{\frac{-j-4}{2}}\, \zeta(j+2) +
R^{\,alt}_{I_{N_j}}\Big) \end{split}\eeq{E0N2B} with \beq
R^{\,alt}_{I_{N_j}}
=\sum_{n=1}^{\infty}\,\sumprime_{\substack{\ell_i=-\infty\\i=1,\ldots,
j}}^{\infty}\dfrac{2\,\,n^{\frac{j+1}{2}}}{\pi}\,\dfrac{\,K_{\frac{j+1}{2}}
\big(\,2\pi\,n\,\sqrt{(\ell_1\frac{a}{a_{m-1}})^2+\cdots+(\ell_j\,\frac{a_{k_j
-1}}{a_{m-1}})^2}\,\,\,\big)}
{\left[(\ell_1\frac{a}{a_{m-1}})^2+\cdots+(\ell_j\frac{a_{k_j-1}}{a_{m-1}})^2\right]^{\tfrac{j+1}{4}}}
\eeq{RNj2B} For Dirichlet given by \reff{E0DD}, the case $j\eq0$
does not yield any $L_1$ terms so that it can be dropped. $L_1=a$
appears only in the numerator via $L_{k_1}=a$ when $k_1=1$ (with the
other lengths given by $L_{k_j}=a_{k_j-1}$). We obtain  \beq
E^{\,alt}_{0_{D_I}}(a)=\dfrac{\pi}{2^{d+1}}\sum_{j=1}^{d-1} (-1
)^{d+j}\,\,\xi^{\,d-1}_{\,1,.., k_j}\,\dfrac{a\,a_{k_2-1}\ldots
a_{k_j-1}}{(a_{d-1})^{j+1}}\Big(
\Gamma(\tfrac{j+2}{2})\,\pi^{\frac{-j-4}{2}}\, \zeta(j+2) +
R^{\,alt}_{I_{\!D_j}}\Big)\eeq{E0D2B} with \beq R^{\,alt}_{I_{D_j}}
=\sum_{n=1}^{\infty}\,\sumprime_{\substack{\ell_i=-\infty\\i=1,\ldots,
j}}^{\infty}\dfrac{2\,\,n^{\frac{j+1}{2}}}{\pi}\,\dfrac{\,K_{\frac{j+1}{2}}
\big(\,2\pi\,n\,\sqrt{(\ell_1\frac{a}{a_{d-1}})^2+\cdots+(\ell_j\,\frac{a_{k_j-1}}{a_{d-1}})^2}\,\,\,\big)}
{\left[(\ell_1\frac{a}{a_{d-1}})^2+\cdots+(\ell_j\frac{a_{k_j-1}}{a_{d-1}})^2\right]^{\tfrac{j+1}{4}}}\,.
\eeq{RDj2B}

The alternative expression for the Casimir force in region I for
Neumann is \beq
\begin{split}F^{\,alt}_{N_I}&= -\dfrac{\partial}{\partial
a}\,E^{\,alt}_{0_{N_I}}(a)\\&=-\dfrac{\pi}{24\,a^2}+\dfrac{\pi}{2^{d+1}}\!\sum_{m=2}^{d}\!\sum_{j=1}^{m-1}
2^{d-m}\,\xi^{\,m-1}_{\,1,k_2,.., k_j}\,\dfrac{a_{_{k_2-1}}\ldots
a_{_{k_j-1}}}{(a _{m-1})^{j+1}}\,
\Gamma(\tfrac{j+2}{2})\,\pi^{\frac{-j-4}{2}}\,
\zeta(j+2)-\dfrac{\partial R^{\,alt}_{I_N}}{\partial a}
\end{split}\eeq{FaltNI}
where  \beq
R^{\,alt}_{I_N}=-\dfrac{\pi}{2^{d+1}}\!\sum_{m=2}^{d}\!\sum_{j=1}^{m-1}
2^{d-m}\,\xi^{\,m-1}_{\,1,k_2,.., k_j}\,\dfrac{a\,a_{_{k_2-1}}\ldots
a_{_{k_j-1}}}{(a _{m-1})^{j+1}}\, R^{\,alt}_{I_{N_j}}\,.\eeq{RINAlt}
The corresponding alternative expression for Dirichlet in region I
is \beq
\begin{split}F^{\,alt}_{D_I}&= -\dfrac{\partial}{\partial
a}\,E^{\,alt}_{0_{D_I}}(a)\\&=-\dfrac{\pi}{2^{d+1}}\sum_{j=1}^{d-1}
(-1 )^{d+j}\,\,\xi^{\,d-1}_{\,1,.., k_j}\,\dfrac{a_{k_2-1}\ldots
a_{k_j-1}}{(a_{d-1})^{j+1}}\,
\Gamma(\tfrac{j+2}{2})\,\pi^{\frac{-j-4}{2}}\, \zeta(j+2)\,
-\dfrac{\partial R^{\,alt}_{I_D}}{\partial a}
\end{split}\eeq{FaltDI}
where \beq R^{\,alt}_{I_D}=\dfrac{\pi}{2^{d+1}}\sum_{j=1}^{d-1} (-1
)^{d+j}\,\,\xi^{\,d-1}_{\,1,.., k_j}\,\dfrac{a\,a_{k_2-1}\ldots
a_{k_j-1}}{(a_{d-1})^{j+1}}\, R^{\,alt}_{I_{\!D_j}}\,.\eeq{RIDAlt}

To obtain the Casimir force on the piston we need to add the
contribution from region II:  $F_{N_{I\!I}}$ and $F_{D_{I\!I}}$
given by \reff{FN2} and \reff{FD2} respectively. For Neumann we
obtain  \beq F^{\,alt}_N = F^{\,alt}_{N_I} + F_{N_{I\!I}}=
-\dfrac{\pi}{24\,a^2}-\dfrac{\partial R^{\,alt}_{I_N}}{\partial a}
-\dfrac{\pi}{2^{d+1}}\!\sum_{m=3}^{d}\!\sum_{j=2}^{m-1}
2^{d-m}\,\xi^{\,m-1}_{\,1,k_2,.., k_j}\,\dfrac{a_{_{k_2-1}}\ldots
a_{_{k_j-1}}}{(a _{m-1})^{j+1}}\,
R_{{I\!I}_{N_j}}\!(\ell_1\!=\!0)\eeq{FaltN} where
$R_{{I\!I}_{N_j}}\!(\ell_1\!=\!0)$ is \reff{RNj2} evaluated at
$\ell_1\eq0$. The above can be reduced to a more compact expression
by noticing that the $\ell_1\eq0$ contribution to $-\partial
R^{\,alt}_{I_N}/\partial a$ cancels out with the last term in
\reff{FaltN}. The alternative expression for the Casimir force on
the piston for Neumann boundary conditions reduces to \beq
F^{\,alt}_N = -\dfrac{\pi}{24\,a^2}-\dfrac{\partial}{\partial a}
 R^{\,alt}_{I_N}(\ell_1\!\ne\!0)\,.\eeq{FaltN22} To evaluate
 \reff{FaltN22} we exclude $\ell_1=0$ in \reff{RINAlt} so that $R^{\,alt}_{I_{N_j}}$
 given by  \reff{RNj2B} is evaluated without including $\ell_1\eq0$
 i.e.\beq
R^{\,alt}_{I_{N_j}}(\ell_1\!\ne\!0)
=\sum_{n=1}^{\infty}\,\sum_{\ell_1=1}^{\infty}\sum_{\substack{\ell_i=-\infty\\i=2,\ldots,
j}}^{\infty}\dfrac{4\,\,n^{\frac{j+1}{2}}}{\pi}\,\dfrac{\,K_{\frac{j+1}{2}}
\big(\,2\pi\,n\,\sqrt{(\ell_1\frac{a}{a_{m-1}})^2+\cdots+(\ell_j\,\frac{a_{k_j
-1}}{a_{m-1}})^2}\,\,\,\big)}
{\left[(\ell_1\frac{a}{a_{m-1}})^2+\cdots+(\ell_j\frac{a_{k_j-1}}{a_{m-1}})^2\right]^{\tfrac{j+1}{4}}}\,.
\eeq{RNj2B2} Compared to \reff{RNj2B}, there is no longer a prime on
the sum over $\ell_i$ and $i$ starts with the integer 2 instead of
1. For Dirichlet one obtains \beq F^{\,alt}_D = F^{\,alt}_{D_I} +
F_{D_{I\!I}}= -\dfrac{\partial R^{\,alt}_{I_D}}{\partial a}
+\dfrac{\pi}{2^{d+1}}\,\sum_{j=2}^{d-1}
(-1)^{d+j}\,\xi^{\,d-1}_{\,1,k_2,.., k_j}\,\dfrac{a_{_{k_2-1}}\ldots
a_{_{k_j-1}}}{(a _{d-1})^{j+1}}\,
R_{{I\!I}_{\!D_j}}\!(\ell_1\!=\!0)\eeq{FaltD} where
$R_{{I\!I}_{D_j}}\!(\ell_1\!=\!0)$ is \reff{RDj2} evaluated at
$\ell_1=0$. The above can also be reduced to a more compact
expression since the $\ell_1\eq0$ contribution of $-\partial
R^{\,alt}_{I_D}/\partial a$ cancels out with the last term in
\reff{FaltD}. The alternative expression for the Casimir force on
the piston for Dirichlet boundary conditions reduces to the simple
expression  \beq F^{\,alt}_D = -\dfrac{\partial
R^{\,alt}_{I_D}}{\partial a}(\ell_1\!\ne\!0) \,.\eeq{FaltD22} To
evaluate \reff{FaltD22} we exclude $\ell_1\eq0$ in \reff{RIDAlt} so
that $R^{\,alt}_{I_{\!D_j}}$ given by \reff{RDj2B} is evaluated
without $\ell_1\eq0$ i.e.
 \beq R^{\,alt}_{I_{D_j}}(\ell_1\!\ne\!0)
=\sum_{n=1}^{\infty}\,\sum_{\ell_1=1}^{\infty}\sum_{\substack{\ell_i=-\infty\\i=2,\ldots,
j}}^{\infty}\dfrac{4\,\,n^{\frac{j+1}{2}}}{\pi}\,\dfrac{\,K_{\frac{j+1}{2}}
\big(\,2\pi\,n\,\sqrt{(\ell_1\frac{a}{a_{d-1}})^2+\cdots+(\ell_j\,\frac{a_{k_j-1}}{a_{d-1}})^2}\,\,\,\big)}
{\left[(\ell_1\frac{a}{a_{d-1}})^2+\cdots+(\ell_j\frac{a_{k_j-1}}{a_{d-1}})^2\right]^{\tfrac{j+1}{4}}}\,.
\eeq{RDj2B2}

Our alternative expression for the Casimir force on the piston for
the Neumann case is $F^{\,alt}_N$ which is given by \reff{FaltN22}
together with \reff{RINAlt} and \reff{RNj2B2}. For the Dirichlet
case the alternative expression is $F^{\,alt}_D$ which is given by
\reff{FaltD22} together with \reff{RIDAlt} and \reff{RDj2B2}. It is
now trivial to see that in the limit as the plate separation $a$
tends to infinity that the Casimir force on the piston is zero since
the modified Bessel functions and their derivatives that appear in
$F^{\,alt}_N$ and $F^{\,alt}_D$ via \reff{RNj2B2} and \reff{RDj2B2}
decrease exponentially fast to zero as $a$ tends to infinity.

\end{appendix}
\section*{Acknowledgments}

This work was supported by the Natural Sciences and Engineering
Research Council of Canada (NSERC).

\end{document}